\documentclass[review]{elsarticle}
\usepackage{graphicx}
\usepackage{lineno,hyperref}
\usepackage{subfigure}
\usepackage{amsmath, amssymb}
\usepackage{amsbsy}
\usepackage{mathptmx}
\usepackage{bm}
\usepackage{makecell}
\usepackage{color}
\usepackage{float}
\usepackage{multirow}
\modulolinenumbers[5]

\journal{Elsevier}

\begin{document}

\begin{frontmatter}

\title{Towards optimal high-order compact schemes for simulating compressible flows}

\author[mymainaddress]{Huaibao Zhang}
\ead{zhanghb28@mail.sysu.edu.cn}

\author[mysecondaryaddress]{Fan Zhang\corref{mycorrespondingauthor}}
\ead{zhangfan3@sysu.edu.cn, a04051127@mail.dlut.edu.cn}
\cortext[mycorrespondingauthor]{Corresponding author.}

\author[mysecondaryaddress]{Chunguang Xu}

\address[mymainaddress]{School of Physics, Sun Yat-sen University, Guangzhou 510006, Guangdong, China }
\address[mysecondaryaddress]{School of Aeronautics and Astronautics, Sun Yat-sen University, Guangzhou 510006, Guangdong, China }

\begin{abstract}
Weighted compact nonlinear schemes (WCNS) [Deng and Zhang, JCP 165(2000): 22-44] were developed to improve the performance of the compact high-order nonlinear schemes (CNS) by utilizing the weighting technique originally designed for WENO schemes, and excellent shock capturing capability and  high resolution are achieved. Various work has been given for further improving the performance of WCNSs since then.
In this work, the ENO-like stencil selection procedure of Targeted ENO schemes [Fu et al. JCP 305(2016):333-359] is introduced for interpolating midpoint variables, targeting compact nonlinear schemes which fully abandon the oscillatory stencils crossing discontinuities, and   directly apply optimal linear weights when the flow field is smooth, such that the optimal numerical resolution is fully recovered in smooth flow field. Several canonical numerical cases of scalar equations and the Euler equations of gas dynamics are given to examine the performance of the presented method.
\end{abstract}

\begin{keyword}
compact nonlinear schemes; ENO-like stencil-selection; optimal linear weights; gas dynamics
\end{keyword}

\end{frontmatter}

\section{Introduction} \label{sec:Intro}

The spatial solution of flow field containing strong discontinuities such
  as shock waves is a challenging topic ever since the shock-capturing schemes invented.
High-order accurate and high-resolution schemes with discontinuity-capturing ability are especially desired in simulating multi-scale compressible flows.
To achieve higher-order accuracy in smooth flow regions, and to suppress spurious oscillations near discontinuities are two essential problems which especially attract the attention of researchers. These two problems, however, somehow contradict with each other, since oscillation usually comes with the use of high-order reconstruction schemes, and high-order schemes owning discontinuity-capturing capability tends to locally degenerate to lower-order schemes for  robustness or  stability purpose.

Great efforts have been made in the development of new high-order discontinuity-capturing schemes to address the dilemma.
Among those schemes, the Weighted nonlinear schemes, specifically Weighted Essentially Non-Oscillatory (WENO) schemes \cite{Liu1994,Jiang1996,Henrick2005,Borges2008} and Weighted Compact Nonlinear Schemes (WCNS) \cite{Deng2000} are two typical classes of methods, which have achieved great success. They can adaptively tune the numerical dissipation by changing the nonlinear convex combination of candidate stencils according to local flow features, such that high order accuracy and non-oscillatory property near discontinuities can be achieved.

Compact finite difference schemes have shown spectral-like resolution \cite{Lele1992}, which is highly favored in the simulations of  multi-scale flow problems. 
In certain circumstances, WCNS also has several advantages over the standard finite-difference WENO: (1) the resolution is slightly higher; (2) various numerical flux schemes can be applied, including Roe's flux difference splitting (FDS) scheme \cite{Roe1981}, van Leer's flux vector splitting  scheme \cite{Leer1982}, and Liou's advection upstream splitting method (AUSM) \cite{Liou1993a}; and (3) WCNS performs well on freestream and vortex preservation properties on wavy grids  \cite{Nonomura2010}.

Therefore,   various researches have been attracted in the improvements and applications of WCNS.
The complete WCNS procedure includes three steps \cite{Deng2000}: (1) node-to-midpoint weighted nonlinear interpolation of given variables, (2) flux evaluation at the midpoint, and (3) midpoint-to-node high-order differencing of flux function.
In the second step, various upwind flux schemes can be applied, and related investigations have been given \cite{Wang2016,TU2016_AMM,ZhuHJ2017}. In the third step, various candidate midpoint-to-node or midpoint-and-node-to-node difference schemes are also available. Deng and Zhang  \cite{Deng2005_AIAA} indicate that the type of midpoint-to-node difference scheme, explicit or implicit, does not affect the resolution of a fifth- or fourth-order WCNS significantly, because the weighted nonlinear interpolation, i.e.  step (1), dominates the resolution property. In addition, Nonomura  et al. \cite{Nonomura2009} demonstrate that the type of midpoint-to-node difference scheme does not significantly change the resolution, even for higher-order WCNSs. Moreover, Nonomura and Fujii \cite{Nonomura2013} proposed an explicit formula which can significantly improve the robustness of WCNS and this formula is also more compact because it is featured by using a midpoint-and-node-to-node difference scheme. In general, aforementioned work suggests that explicit difference scheme is preferred, due to satisfying resolution with lower cost and simplicity for parallelization and vectorization.

Resolutions of WCNSs have been greatly improved by focusing on the improvement of its first construction step, i.e. node-to-midpoint weighted nonlinear interpolation. Novel nonlinear weights achieving optimal order of accuracy \cite{Henrick2005,Borges2008} are introduced into WCNSs \cite{Yan2016,Yan2017}, and novel nonlinear compact node-to-midpoint interpolation is also implemented \cite{Ma_CJCM_2015}. Recently, a family of high-order targeted ENO schemes has been proposed \cite{Fu2016}. Apart from using novel nonlinear weights and incremental-width candidate stencils, the TENO scheme is characterized by using a newly developed ENO-like stencil-selection procedure. These features of TENO schemes bring significant improvement in spatial resolution. Particularly, the ENO-like stencil-selection procedure is essential to fully recover the background linear schemes in smooth region.
Further developments and applications of TENO schemes can be found in \cite{Fu2017,Haimovich2017,FU_CICP2018,Fu2018}, and boundary variation diminishing (BVD) method was also used to sharpen the captured shock waves with using TENO scheme \cite{Sun2016a}. In this work, which is the further development of compact nonlinear schemes \cite{Deng1997},
the specific ENO-like stencil-selection procedure is adopted with an aim that the optimal node-to-midpoint interpolation of compact linear scheme is achieved in smooth fields, including in the region of smooth critical points.


 This article is organized as follows. In section \ref{sec:method}, the WCNS scheme and the presented method, labeled as TCNS similarly to the nomenclature of TENO,
 are introduced in detail. In section \ref{sec:results}, detailed numerical analysis including ADR (Approximate dispersion relation) analysis \cite{Pirozzoli2006} and the solutions of scalar equation and system equations, are introduced to testify the performance of the presented method. Finally, concluding remarks are given in the last section.

\section{Numerical methods} \label{sec:method}

The governing equations of compressible flows are  hyperbolic systems. Without loss of generality, the theoretical analysis and numerical solutions of the one-dimensional scalar hyperbolic conservation law
can be first used to examine the performance of numerical schemes, and then the associated results can be extended to one- or two-dimensional hyperbolic system of equations without substantial difficulty.

The one-dimensional hyperbolic conservation law can be written as
\begin{equation}\label{eq:hcl}
\frac{{\partial u}}{{\partial t}} + \frac{{\partial f(u)}}{{\partial x}} =0 ,
\end{equation}

\noindent in which the characteristic velocity is $\frac{{\partial f(u)}}{{\partial u}}$ and assumed to be positive, without loss of generality. Here, the spatial discretization of Eq.~\eqref{eq:hcl}  is given on an  equally spaced one-dimensional mesh, in which the distance between two grid nodes are $h$, leading to an ordinary differential equation (ODE) system, i.e.

\begin{equation}\label{eq:dis}
\frac{{d u_i}}{{d t}}=- \frac{{\partial f}}{{\partial x}}|_{x=x_i}=-f_i', \quad i=1, \cdots, n.
\end{equation}
\noindent The numerical  solutions of $f_i'$ will be discussed in the following sections, and the temporal solutions are given by using the third-order strongly  stable  Runge-Kutta  method \cite{Gottlieb2001}.

\subsection{The approximation of the flux derivative}

The first-order derivative of flux function, i.e. $f_i'$, can be approximated by using various numerical schemes. In a WCNS-type method, usually a central (compact) difference scheme is used to calculate the flux derivatives at grid nodes, by utilizing the flux function at midpoints and grid nodes.

As aforementioned, because of their efficiency, in this work, explicit central difference schemes which require nontridiagonal inversion are used to approximate the derivative. Here, for achieving fifth-order overall accuracy, three midpoint-to-node (MD) or midpoint-and-node-to-node (MND) formulas of sixth-order accuracy are given as

\begin{equation} \label{eq:wcns}
  f_{i}'=\frac{a_1}{h}(\widetilde{f}_{i+\frac{1}{2}}-\widetilde{f}_{i-\frac{1}{2}})+\frac{a_2}{h}(\widetilde{f}_{i+\frac{3}{2}}-\widetilde{f}_{i-\frac{3}{2}})+\frac{a_3}{h}(\widetilde{f}_{i+\frac{5}{2}}-\widetilde{f}_{i-\frac{5}{2}}),
\end{equation}

\begin{equation} \label{eq:hwcns}
  f_{i}'=\frac{b_1}{h}(\widetilde{f}_{i+\frac{1}{2}}-\widetilde{f}_{i-\frac{1}{2}})+\frac{b_2}{h}(\widetilde{f}_{i+1}-\widetilde{f}_{i-1})+\frac{b_3}{h}(\widetilde{f}_{i+2}-\widetilde{f}_{i-2}),
\end{equation}

\noindent   and

\begin{equation} \label{eq:mnd}
  f_{i}'=\frac{c_1}{h}(\widetilde{f}_{i+\frac{1}{2}}-\widetilde{f}_{i-\frac{1}{2}})+\frac{c_2}{h}(\widetilde{f}_{i+1}-\widetilde{f}_{i-1})+\frac{c_3}{h}(\widetilde{f}_{i+\frac{3}{2}}-\widetilde{f}_{i-\frac{3}{2}}),
\end{equation}

\noindent  which are corresponding to the WCNS \cite{Deng2002}, Hybrid cell-edge and cell-node Weighted Compact Nonlinear Scheme (HWCNS) \cite{Deng2011} and WCNS-midpoint-and-node-to-node difference (WCNS-MND) \cite{Nonomura2013}, respectively.  Moreover, these three formulas show that they are not compact schemes, but can be taken as the special cases of the general WCNS \cite{Deng2002}. The constant coefficients in Eq.~\eqref{eq:wcns},~\eqref{eq:hwcns} and \eqref{eq:mnd}, can be found in the corresponding references. 


It is worth noticed that the required midpoint stencils of HWCNS and WCNS-MND are more compact than that of Eq.(\ref{eq:wcns}), because  they use node variables in the difference schemes, involving fewer midpoint interpolations. Moreover,
Nonomura and Fujii \cite{Nonomura2013} proved that WCNS-MND is more robust but less accurate than the original WCNS \cite{Deng2000}and the original WCNS has higher propensity to blow up compared to WENO schemes when strong discontinuities are captured. 
WCNS-MND method is thus used in simulations of turbulence problems \cite{Zhao2018} and multi-species flow problems \cite{Wong2017}.
In this work, most of the numerical results and discussions are given by using the sixth-order explicit MD scheme (Eq.~\eqref{eq:wcns}), and, of course, the other two formulas, Eq.~\eqref{eq:hwcns} and \eqref{eq:mnd}, can  be used straightforwardly.


\subsection{Node-to-midpoint interpolation: WCNS}

As shown in the last subsection, midpoint flux terms are unknown and should be evaluated before performing the MD or MND procedure. To achieve this goal, numerical
upwind flux functions are used. Without loss of generality, a scalar form is used to explain the interpolation procedure. The scalar upwind flux function is
\begin{equation}
f_{i\pm\frac{1}{2}}=\frac{1}{2}\left[\left(f(u_{R,i\pm\frac{1}{2}})+f(u_{L,i\pm\frac{1}{2}})\right)-|\hat{a}|\left(u_{R,i\pm\frac{1}{2}}-u_{L,i\pm\frac{1}{2}}\right)\right],
\end{equation}
\noindent where the subscript $L$ and $R$ indicate the variables at the left and right side of $x_{i\pm\frac{1}{2}}$, respectively, and $\hat{a}$ is the approximate eigenvalue.
The question left to be settled is how to calculate $u_{L/R,i\pm\frac{1}{2}}$ accurately, while maintaining numerical stability, nonoscillatory and sharp discontinuity-capturing properties. For simplicity, we only consider the evaluation of variables on the left   side of $x_{i+\frac{1}{2}}$, i.e. $u_{L,i+\frac{1}{2}}$, in the following paragraphs. The interpolations of  the other midpoint variables are performed by using a symmetrical form of $u_{L,i+\frac{1}{2}}$.

The fifth-order node-to-midpoint reconstruction of WCNS to be used in this work was introduced in \cite{Deng2000}. The basic idea of it is to approximate a linear optimal approximation of the midpoint variable
\begin{equation}
u_{L,i+\frac{1}{2}}=u_i+\frac{1}{128}\left(3u_{i-2}-20u_{i-1}-38u_{i}+60u_{i+1}-5u_{i+2}\right) ,
\end{equation}
by using a nonlinear convex combination of lower-order interpolations.
It is obvious that a five-point full stencil $S_{i+\frac{1}{2}}=\{x_{i-2},x_{i-1},x_{i},x_{i+1},x_{i+2}\}$ is used to calculate the high-order approximation.
This optimal fifth-order scheme can be equivalently represented by using three third-order polynomials  each constructed on the following three-point substencil
\begin{equation}
S_{i+\frac{1}{2},k}=\{x_{i+k-3},x_{i+k-2},x_{i+k-1}\}, \quad k=1,2,3.
\end{equation}

Each of the third-order polynomials can be expressed in a generic form using the (approximated) $n-$th derivatives ($n=1,2$)
\begin{equation}
u_{L,i+\frac{1}{2},k}=u_i\left(x_i+{\Delta x}\right)=u_i+ u^{(1)}_{i,k}\Delta x+u^{(2)}_{i,k}\frac{\Delta x^2}{2},
\end{equation}
\noindent where $\Delta x=x_{i+\frac{1}{2}}-x_i=\frac{h}{2}$. Specifically, the first- and second-order derivatives are
\begin{equation}
\begin{split}
& u^{(1)}_{i,1}=\frac{1}{2h}(u_{i-2}-4u_{i-1}+3u_i),\\
& u^{(1)}_{i,2}=\frac{1}{2h}(u_{i+1}-u_{i-1}),\\
& u^{(1)}_{i,3}=\frac{1}{2h}(-3u_{i}+4u_{i+1}-u_{i+2}),
\end{split}
\end{equation}
\noindent and
\begin{equation}
\begin{split}
& u^{(2)}_{i,1}=\frac{1}{h^2}(u_{i-2}-2u_{i-1}+u_i),\\
& u^{(2)}_{i,2}=\frac{1}{h^2}(u_{i-1}-2u_{i}+u_{i+1}),\\
& u^{(2)}_{i,3}=\frac{1}{h^2}(u_{i}-2u_{i+1}+u_{i+2}),
\end{split}
\end{equation}
\noindent respectively. The linear optimal scheme is then represented as
\begin{equation}
u_{L,i+\frac{1}{2}}=   \sum\limits_{k=1}^3 d_k u_{L,i+\frac{1}{2},k},
\end{equation}
\noindent where the optimal linear weights are
\begin{equation}
 d_1=\frac{1}{16}, \quad d_2=\frac{10}{16}, \quad d_3=\frac{5}{16}.
\end{equation}

Obviously, directly using the optimal weights yields a fifth-order accurate scheme. The resulting scheme, however, is inadequate to obtain a satisfying resolution for a steep gradient solution: it causes spurious oscillations. The nonlinear weights of Jiang and Shu \cite{Jiang1996} can be applied to replace the optimal linear weights in order to suppress  non-physical oscillations.
The so called JS nonlinear weights are defined by
\begin{equation}
\omega_k=\frac{\alpha_k}{\sum_{k=1}^3\alpha_k}, \quad \alpha_k=\frac{d_k}{(\beta_k+\epsilon)^2},
\end{equation}
\noindent where the small parameter $\epsilon=10^{-6}$ is specified to avoid the denominator becoming zero, and $\beta_k$ is the smoothness indicator, which is defined as
\begin{equation}
\beta_k=\left(hu^{(1)}_{i,k}\right)^2+\left(h^2u^{(2)}_{i,k}\right)^2.
\end{equation}
\noindent
The main advantage of using nonlinear weights is that the contribution of oscillatory stencil(s) will be approximately eliminated in the final interpolation while preserving those of relative smooth, and thus the spurious numerical oscillation can be suppressed.

\subsection{Node-to-midpoint interpolation: ENO-like stencil-selection} \label{sec:tcns}

Instead of merely concentrating on improving the  nonlinear weights, the ENO-like stencil-selection procedure  \cite{Fu2016} is introduced as an essential component of the presented method.
Firstly, the nonlinear smoothness measurement yielding strong scale-separation mechanism of the fifth-order scheme is given as
\begin{equation} \label{eq:TENO1}
\gamma_k=\left(C+\frac{\tau_5}{\beta_k+\epsilon} \right)^q, \quad k=  1, 2, 3.
\end{equation}
\noindent where $\tau_5=|\beta_1-\beta_3|$ is the global smooth indicator that was originally introduced in \cite{Borges2008}, and the small threshold is defined as $\epsilon=10^{-40}$. Constant $C=1$ is set, and the integer power  $q=6$ is used. As introduced by Fu et al. \cite{Fu2016}, larger integer power exponent $q$ and smaller $C$ are preferable for a stronger separation between resolved and non-resolved scales, and the discontinuity-detection capability can be significantly enhanced.

Rather than using the original nonlinear smoothness measurement, Eq.~\eqref{eq:TENO1} is further normalized by
 \begin{equation} \label{eq:norm}
\chi_k=\frac{\gamma_k}{\sum_{k=1}^{3}\gamma_k},
\end{equation}
which is then subject to a sharp cutoff function
\begin{equation}
\delta_k=
\begin{cases}
\begin{matrix}
0, & \text{if} \quad \chi_k < C_T, \\
1, & \text{otherwise}.
\end{matrix}
\end{cases}
\end{equation}
The underlying idea of this design is that by introducing a value $C_T$ as the threshold of smoothness, the candidate stencils are attributed as ``smooth'' or ``oscillatory'', such that those stencils of oscillations are abandoned thereby, and only smooth ones are saved with their associated optimal linear weights used in the final interpolation. The resulting weight functions are now given by
\begin{equation} \label{eq:TENO_W}
\omega_k^{(T)}=\frac{d_k\delta_k}{\sum_{k=1}^{3}d_k\delta_k},
\end{equation}
where $d_k$ denotes the optimal linear weight, and the above cut-off function $\delta_k$ is incorporated into the final weighting to decide whether each candidate stencil is taken into account or not. Two promising advantages of this weighting procedure can be readily shown. Firstly, by removing the stencil crossing discontinuity completely, it ensures the numerical robustness of the scheme. Secondly, the background fifth-order linear scheme can be fully recovered in smooth regions, including at smooth critical points.

\begin{figure}
 \centering
 \includegraphics[width=6cm]{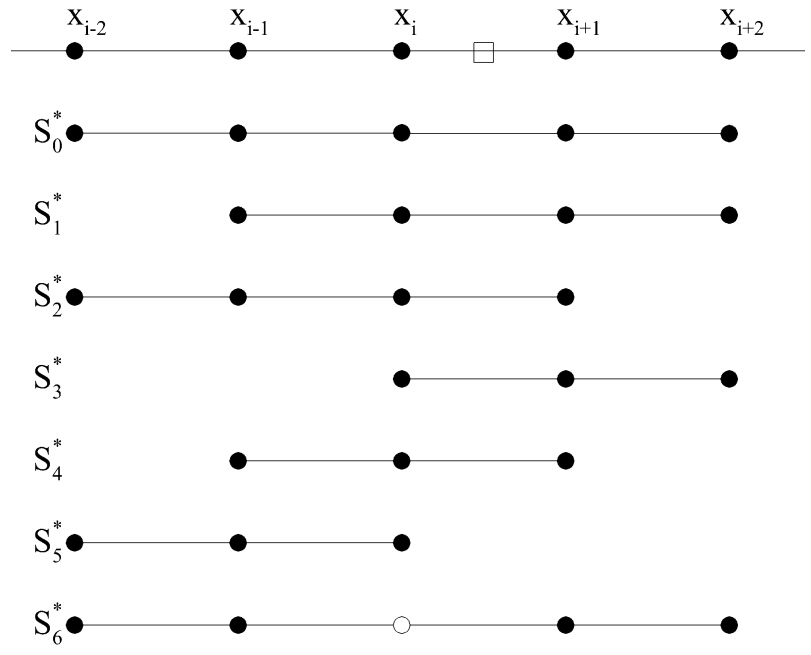}
 \caption{ \label{fig:f:re}
  Schematic of the equivalent candidate stencils of the fifth-order nonlinear interpolation. The circle indicates that the stencil is not \textit{really} continuous.}
\end{figure}

In fact, the stencil-selection procedure has simplified the nonlinear convex combinations of  candidate stencils. Instead of using the continuous varying weights leading to infinite possible combinations, the presented method in fact uses only several candidate combinations, since each candidate stencil has only two possible results of $d_k\delta_k$, i.e. $d_k$ and $0$. Therefore, similar to \cite{Fu2018}, each of the possible combinations of the nonlinear interpolation of $\hat{u}_{L,i+\frac{1}{2}}$ can be represented as $\hat{u}_{L,i+\frac{1}{2},m}^*=   a_{m,i-2}u_{i-2}+a_{m,i-1}u_{i-1} +a_{m,i}u_{i}+a_{m,i+1}u_{i+1}+a_{m,i+2}u_{i+2}$, where the coefficients are given in Table \ref{table:co1}.
 The corresponding equivalent stencil of   each convex combination is given in Fig.\ref{fig:f:re}, by which the meaning of "stencil-selection" can be interpreted more clear.
\begin{table}
\scriptsize
\centering
\caption{ The coefficients of the equivalent single polynomial spatial reconstructions.}\label{table:co1}
\begin{tabular}{cccccccc}
\hline
\text{if} $\delta_{1,2,3}=$ &$\hat{u}_{L,i+\frac{1}{2},m}^*$& S$_m^*$  & $a_{m,i-2}$  &  $a_{m,i-1}$ & $a_{m,i}$ &  $a_{m,i+1}$ &  $a_{m,i+2}$    \\ \hline
1,1,1                       &$\hat{u}_{L,i+\frac{1}{2},0}^*$& S$_0^*$  & 3/128         &  -5/32    &    45/64   & 15/32 &  -5/128     \\
0,1,1                       &$\hat{u}_{L,i+\frac{1}{2},1}^*$& S$_1^*$  & 0            & -1/12       &  5/8       & 1/2  &   -1/24     \\
1,1,0                       &$\hat{u}_{L,i+\frac{1}{2},2}^*$& S$_2^*$  & 3/88         & -5/22     &   75/88    & 15/44  & 0                \\
0,0,1                       &$\hat{u}_{L,i+\frac{1}{2},3}^*$& S$_3^*$  & 0            & 0          &  3/8       & 3/4  & -1/8      \\
0,1,0                       &$\hat{u}_{L,i+\frac{1}{2},4}^*$& S$_4^*$  & 0            &  -1/8      &3/4       & 3/8  &   0   \\
1,0,0                       &$\hat{u}_{L,i+\frac{1}{2},5}^*$& S$_5^*$  & 3/8          & -5/4       &  15/8      & 0    & 0      \\
1,0,1                       &$\hat{u}_{L,i+\frac{1}{2},6}^*$& S$_6^*$  &    1/16      & -5/24      &  5/8     & 5/8  & -5/48    \\
\hline
\end{tabular}
\end{table}

In addition, using the stencil-selection procedure results into several potential choices of high-order polynomials, and each can be optimized independently such as in \cite{Fu2018}.
Parameter $C_T$, serving as the global reference of smooth indicators, is an effective and a direct mean to control the spectral properties of the scheme for compressible turbulence simulation in which close embedded shocklets need to be captured \cite{Fu2018}. Within the scope of this work, however, the parameter is simply set as $C_T=10^{-5}$ for all the simulations without further investigation on the influence of its choice, but the advantages of importing the ENO-like stencil-selection procedure into compact nonlinear schemes are  clearly showed. 

\section{Numerical results} \label{sec:results}

A variety of canonical problems are simulated to assess  fifth-order  WCNS-JS, WCNS-Z, and the proposed scheme TCNS. One-dimensional linear advection equation, one-dimensional inviscid Burgers equation and Euler equations of gas dynamics are used as model equations. The ideal-gas equation of state is given by $p = (\gamma -1 ) \rho e$ with $\gamma = 1.4 $ to close Euler equations. Node-to-midpoint interpolation is performed on characteristic variables to alleviate spurious   oscillations \cite{Deng2000}.
Van leer scheme \cite{Leer1982} is used for the computation of fluxes.  The proposed fifth-order TCNS scheme   is first assessed through comparisons with the classical WCNS-JS, WCNS-Z on spectral properties and official order of accuracy. Its rate of convergence is then verified by solving one-dimensional linear advection equation, and one-dimensional inviscid Burgers equation. Performances of the above fifth-order schemes in shock-capturing and wave resolutions are compared by solving selected test-problems both in one and two dimensions.
The CFL$=0.6$ is used as default for all numerical schemes and testing cases.


\subsection{Approximate dispersion relation analysis}
\begin{figure}[h!t]
\begin{center}
\subfigure[\label{fig:f:stencils2}{}]{
\resizebox*{5.5cm}{!}{\includegraphics{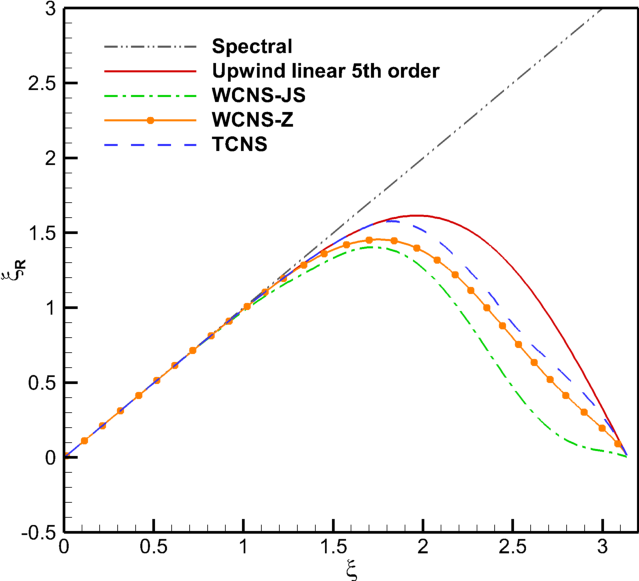}}}
\subfigure[\label{fig:f:stencils1}{}]{
\resizebox*{5.5cm}{!}{\includegraphics{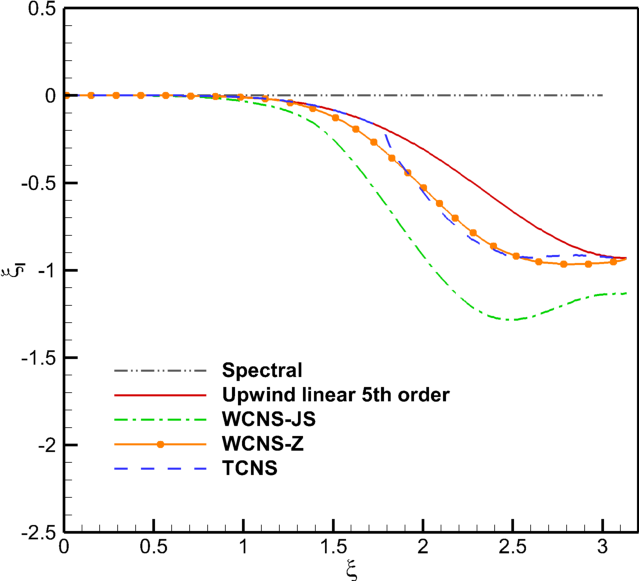}}}
\caption{\label{fig:f:5order} Approximated dispersion and dissipation properties of different fifth-order schemes: dispersion (left) and dissipation (right). }
\end{center}
\end{figure}

Following the ADR analysis introduced by Pirozzoli \cite{2006JCoPh} and Tu et al.~\cite{Tu2012AAS}, the spectral properties of different fifth-order schemes are compared and shown in Fig.~\ref{fig:f:5order}. The result of the presented method agrees very well with the background linear scheme for the low range of wave-numbers with a recovered wave-number reaching up to 1.76. A significant improvement can also be found when cosmpared against the WCNS-JS in both dispersion and dissipation properties. While slight inferior result of dissipation property is given by TCNS when compared with WCNS-Z for the wave-number in intermediate range, TCNS performs better than WCNS-Z scheme for rest range of wave-numbers. Moreover, the performance of TCNS will be significantly improved if small $C_T$ is used, and the adaptive dissipation control in \cite{Fu2018} will further improve its overall performance.

\subsection{Linear advection problem}
The one-dimensional Gaussian pulse advection problem~\cite{Yamaleev2009} is used to assess the numerical order of accuracy of the proposed scheme. This problem is modeled by means of the linear advection equation, given by
\begin{equation}
\frac{\partial u}{\partial t} + \frac{\partial u}{\partial x} = 0, \quad x\in [0,1],
\end{equation}
with periodic boundary-conditions and the initial conditions given by
\begin{equation}
\begin{aligned}
u(t,x=0) &=  u(t,x=1), \\
u(0,x)   &= e^{-300(x-x_c)^2}, \quad x_c = 0.5.
\end{aligned}
\end{equation}

Time integration is performed up to $t=1$, which corresponds to one period of the single wave propagation in time. A set of evenly distributed grids are progressively refined by a factor of 2 from the most coarse grid with $N=51$. Numerical simulation on each of the grid is conducted by using a time step size that even further reducing its value will not change the evaluated  error of the numerical solution.

\begin{figure}[h!t]
\begin{center}
\subfigure[\label{fig:f:LinearAdvection1}{$L_2 $ error}]{
\resizebox*{5.5cm}{!}{\includegraphics{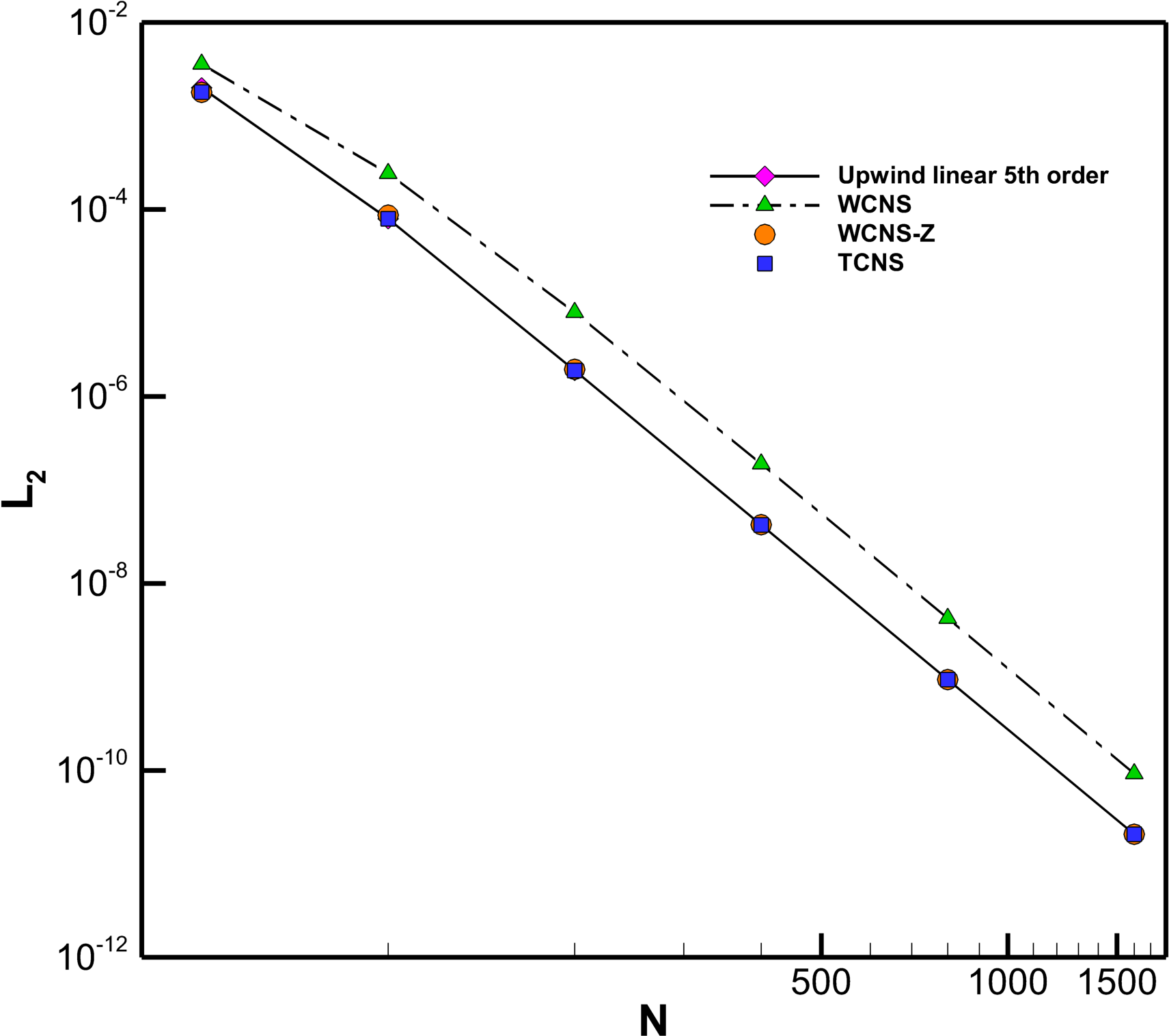}}}
\subfigure[\label{fig:f:LinearAdvection2}{$L_\infty $ error}]{
\resizebox*{5.5cm}{!}{\includegraphics{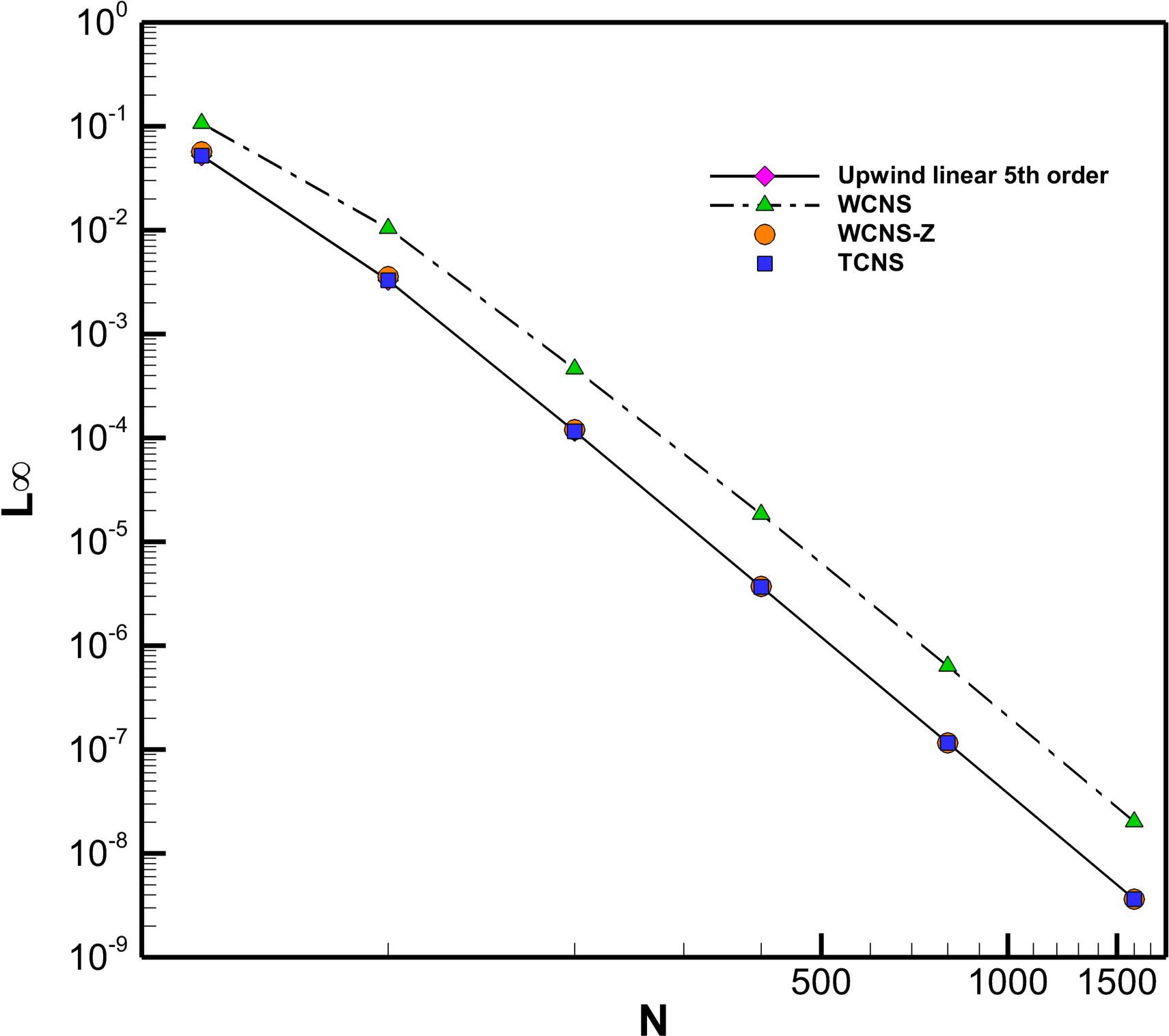}}}
\caption{\label{fig:f:LinearAdvection} Convergence rate of the $L_2 $- and $L_\infty $-error for different fifth-order schemes used by solving the linear advection equation at $t=1$. }
\end{center}
\end{figure}

Table~\ref{linear-table1} and Fig.~\ref{fig:f:LinearAdvection} illustrate the numerical error of those fifth-order schemes used, where the result of TCNS scheme coincides with that of the corresponding linear scheme, indicating that it recovers the optimal order of convergence. A slight deviation can be found for WCNS-Z over the linear scheme for the coarse grids. WCNS-JS also
 shows approximate fifth-order accuracy, but its resolution is significantly lower than WCNS-Z and TCNS. In general, using the ENO-like stencil-selection procedure recovers the optimal linear scheme in this smooth field.
%

\begin{table}[h!t]
\caption{ $L_\infty $-error and convergence rate for different fifth-order schemes used by solving the linear advection equation at $t = 1$ .}
\label{linear-table1}
\begin{tabular}{ccccccccc}
\hline
\multirow{2}{*}{N} & \multicolumn{2}{c}{Linear} & \multicolumn{2}{c}{WCNS-JS} & \multicolumn{2}{c}{WCNS-Z} & \multicolumn{2}{c}{TCNS} \\ \cline{2-9}
 & Error & Order & Error & Order & Error & Order & Error & Order \\
\hline
51 & 5.22E-02 & * & 1.07E-01 & * & 5.67E-02 & * & 5.20E-02 & * \\
101 & 3.30E-03 & 3.98 & 1.04E-02 & 3.37 & 3.57E-03 & 3.99 & 3.30E-03 & 3.98 \\
201 & 1.16E-04 & 4.83 & 4.63E-04 & 4.49 & 1.20E-04 & 4.90 & 1.16E-04 & 4.83 \\
401 & 3.69E-06 & 4.97 & 1.84E-05 & 4.66 & 3.72E-06 & 5.01 & 3.69E-06 & 4.97 \\
801 & 1.16E-07 & 4.99 & 6.36E-07 & 4.85 & 1.16E-07 & 5.00 & 1.16E-07 & 4.99 \\
1601 & 3.64E-09 & 4.99 & 2.02E-08 & 4.98 & 3.64E-09 & 4.99 & 3.64E-09 & 4.99 \\
\hline
\end{tabular}
\end{table}

\subsection{Inviscid Burgers equation}
The one-dimensional inviscid Burgers equation~\cite{HARTEN19973,GEROLYMOS20098481} is used to assess the actual order of accuracy of the proposed scheme when it is applied to a non-linear scalar equation. The governing equation is in the form of
\begin{equation}
\frac{\partial u}{\partial t} + \frac{\partial }{\partial x}\left(\frac{1}{2} u^2\right) = 0, \quad x\in [0,2],
\end{equation}
with periodic boundary-conditions and the initial conditions given by
\begin{equation}
\begin{aligned}
u(t,x=0) &=  u(t,x=1), \\
u(0,x)   &= \frac{1}{2} + \text{sin}(\pi x).
\end{aligned}
\end{equation}

\begin{figure}[h!t]
\begin{center}
\subfigure[\label{fig:burgers1}{$t=0.2$}]{
\resizebox*{5.5cm}{!}{\includegraphics{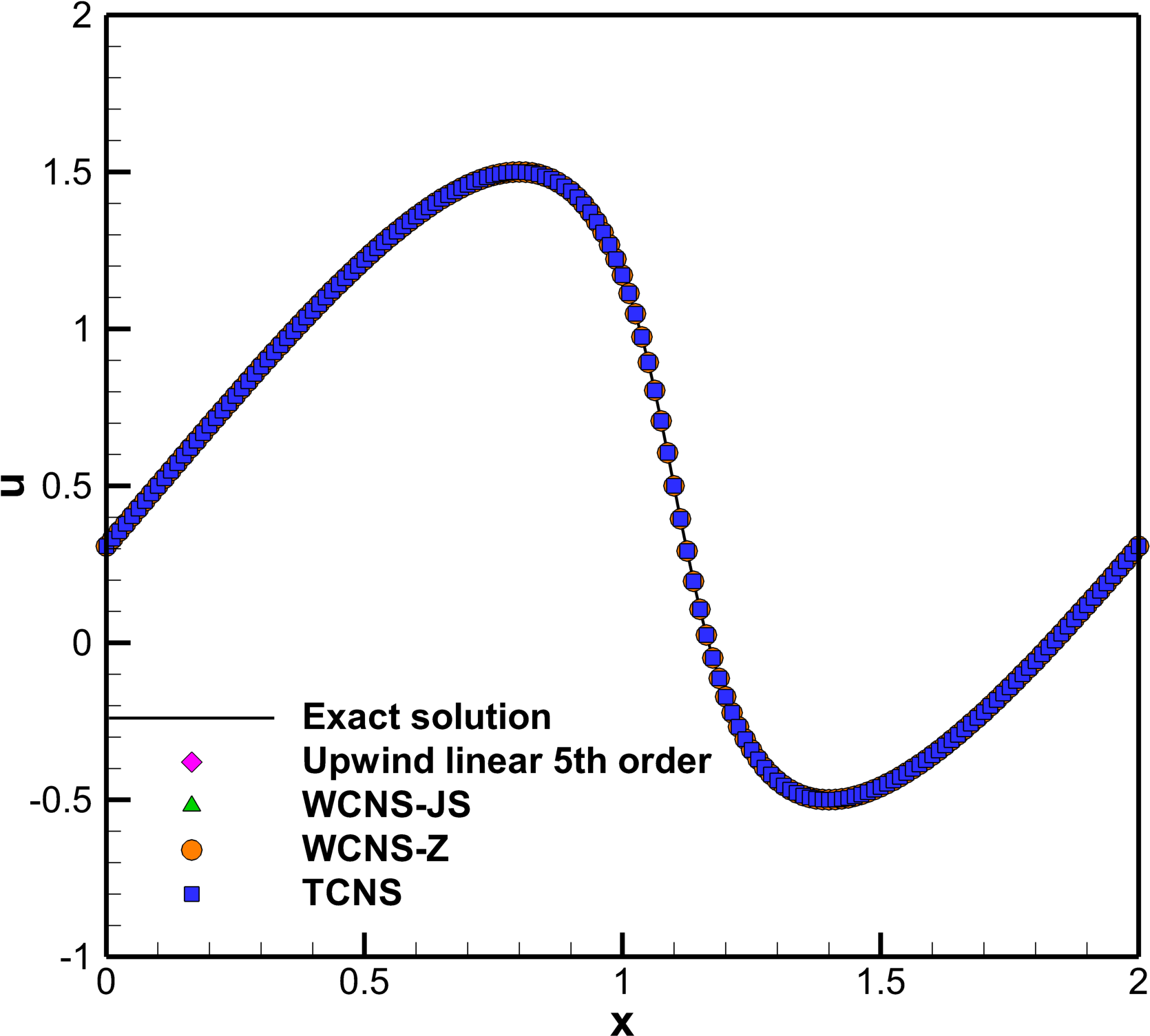}}}
\subfigure[\label{fig:burgers2}{$t=0.7$}]{
\resizebox*{5.5cm}{!}{\includegraphics{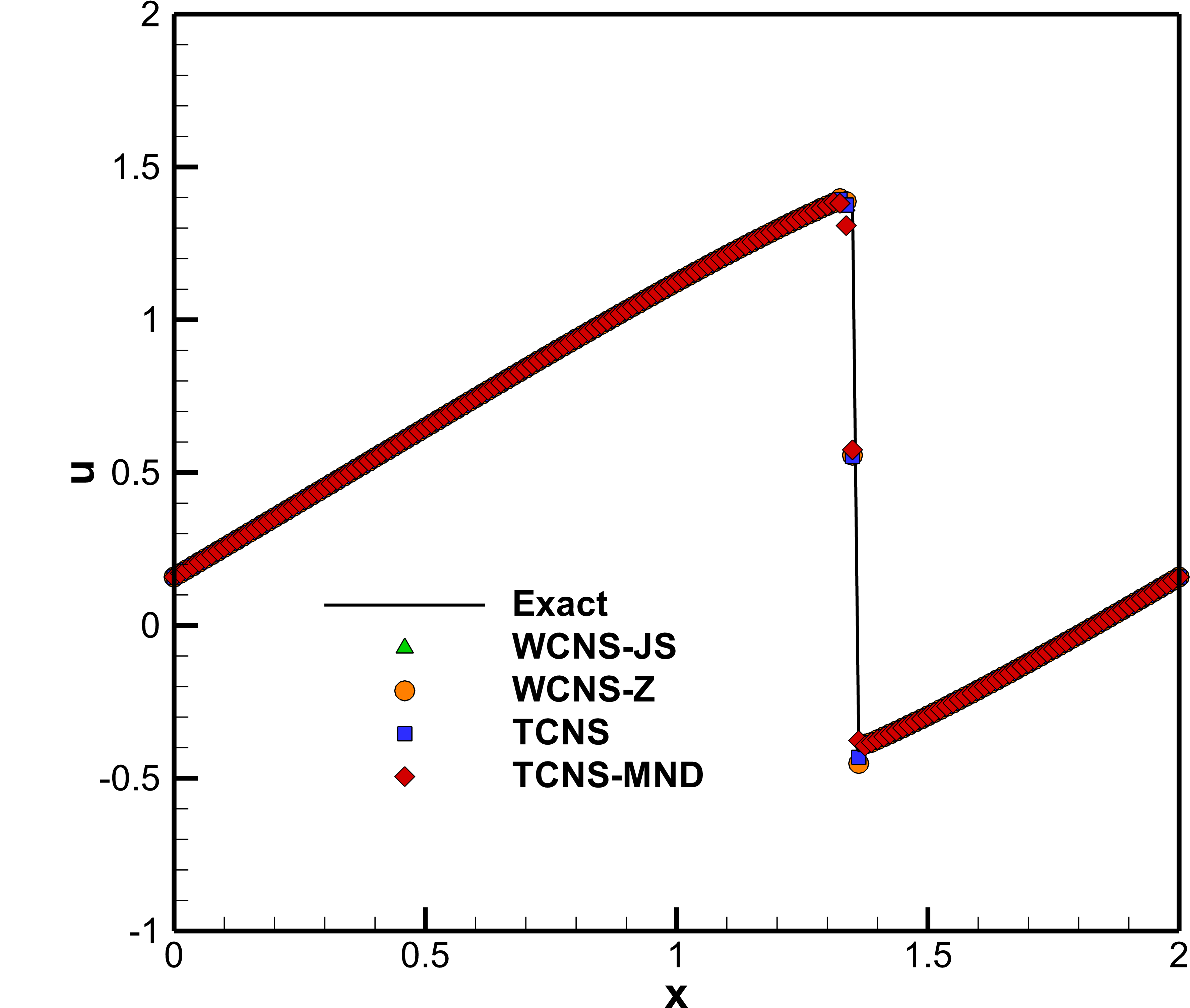}}}
\caption{\label{fig:burgers} One-dimensional inviscid Burgers equation: numerical solutions and exact solution at $t=0.2$ (left) and $t=0.7$ (right). Grid resolution 161. }
\end{center}
\end{figure}

Exact solutions are computed by solving the derived general characteristic relation in reference~\cite{HARTEN19973}. The solutions are smooth for $0 \le t< 1/{\pi}$, and a discontinuity develops and starts to interact with the expansion wave if  $ t \ge 1/{\pi}$. The results at $t=0.2$ and $t=0.7$ are both given to show the continuous and discontinuous distributions. Solutions obtained by the above fifth-order schemes are compared against the exact solution in Fig.~\ref{fig:burgers}, in which, good agreement is presented both for the smooth and the discontinuous solution.
Moreover, at $t=0.2$, $L_\infty $-error and convergence rate for each   scheme  are presented in Table~\ref{table2}.
All of the schemes can achieve fifth-order of accuracy as the grids are refined. Perfect agreement can be found between the TCNS and the background fifth-order linear scheme. WCNS-Z scheme shows minor deviation compared with TCNS scheme only when the grid is relatively coarse.
WCNS-JS scheme is less accurate than the other fifth-order schemes.
Again, TCNS scheme recovers the optimal linear scheme if the solution is smooth.

\begin{table}[h!t]
\caption{ $L_\infty $-error and convergence rate for different fifth-order schemes used by solving the 1-D inviscid Burgers equation at t = 0.2.}
\label{table2}
\begin{tabular}{ccccccccc}
\hline
\multirow{2}{*}{N} & \multicolumn{2}{c}{Linear} & \multicolumn{2}{c}{WCNS-JS} & \multicolumn{2}{c}{WCNS-Z} & \multicolumn{2}{c}{TCNS} \\ \cline{2-9}
 & Error & Order & Error & Order & Error & Order & Error & Order \\ \cline{1-1}
 \hline
41 & 9.96E-04 & * & 1.41E-03 & * & 1.01E-03 & * & 9.96E-04 & * \\
81 & 9.04E-05 & 3.46 & 1.34E-04 & 3.40 & 9.07E-05 & 3.48 & 9.04E-05 & 3.46 \\
161 & 3.21E-06 & 4.81 & 5.06E-06 & 4.73 & 3.22E-06 & 4.82 & 3.21E-06 & 4.81 \\
321 & 1.00E-07 & 5.00 & 1.63E-07 & 4.96 & 1.00E-07 & 5.00 & 1.00E-07 & 5.00 \\
641 & 2.82E-09 & 5.15 & 4.82E-09 & 5.07 & 2.82E-09 & 5.15 & 2.82E-09 & 5.15 \\ \hline
\end{tabular}
\end{table}

Moreover, it should be noticed that in the discontinuous solution, minor overshoot has been found. This problem can be avoided by using the MND method, as shown in Fig.\ref{fig:burgers2}. At the meantime, MND method reduces the resolution of the result, as already introduced by Nonomura and Fujii \cite{Nonomura2013}. Whereas, in this work, only the node-to-midpoint interpolation methods are discussed, and thus the detail of MND method is not further investigated. As mentioned above, the presented method can be applied with using all the midpoint-to-node
 or midpoint-and-node-to-node methods.

\subsection{Sod and Lax shock tube problem}
Riemann initial value problems of Sod~\cite{SOD19781} and Lax~\cite{Lax} are used to evaluate the shock-capturing capability of the proposed scheme. The initial conditions for the Sod problem is
\begin{equation}
(\rho, u, p)=
\begin{cases}
\begin{matrix}
 (1,0,1)        &  \quad  x\in [0,0.5], \\
 (0.125,0,0.1)  & \quad  x\in  \left(0.5,1\right].
\end{matrix}
\end{cases}
\end{equation}
and the final time is $t=2$ by solving the problem on an evenly distributed grid of $N=101$ points.

\begin{figure}[h!t]
\begin{center}
\subfigure[\label{fig:sod-density1}{}]{
\resizebox*{5.5cm}{!}{\includegraphics{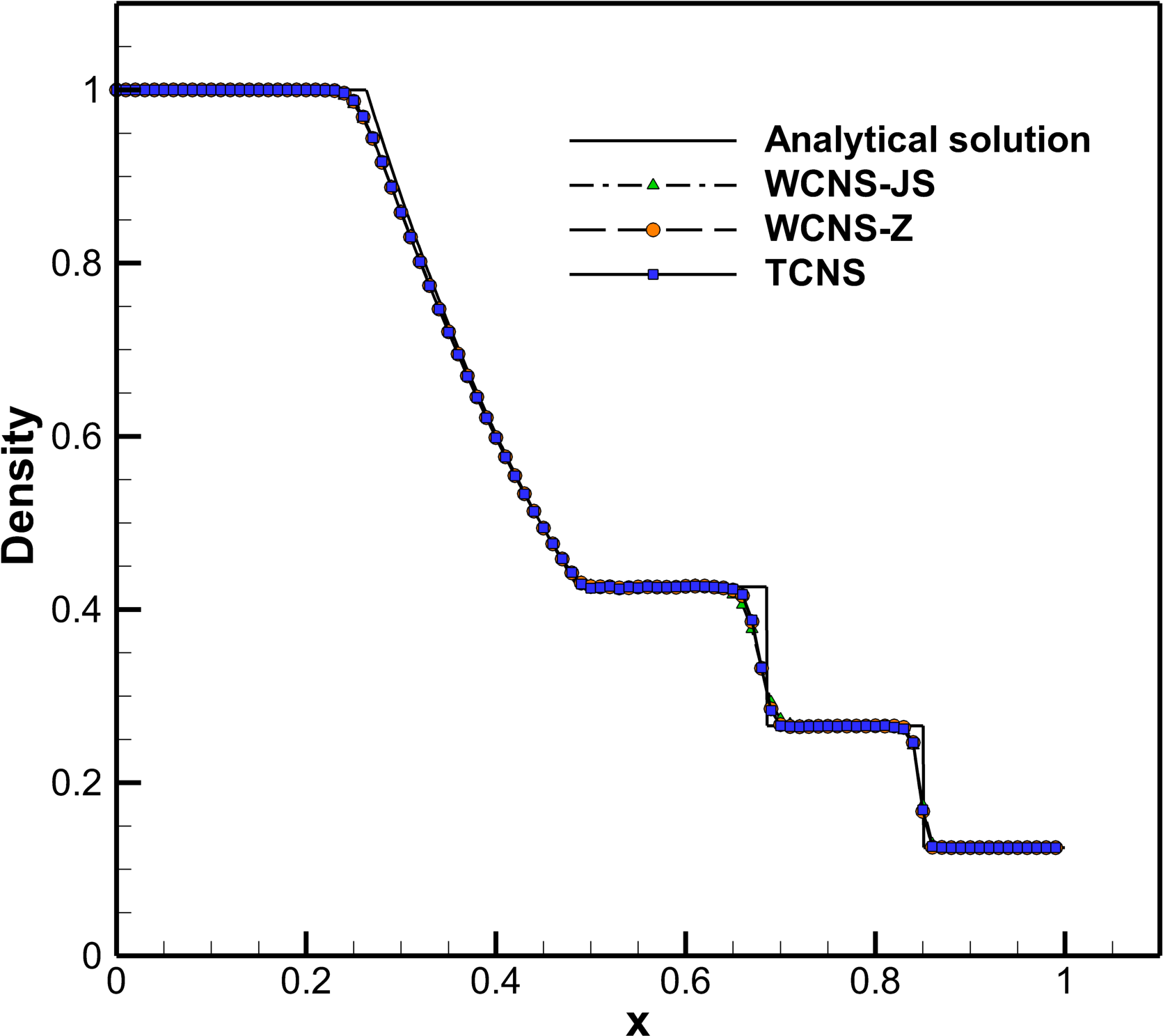}}}
\subfigure[\label{fig:sod-density2}{}]{
\resizebox*{5.5cm}{!}{\includegraphics{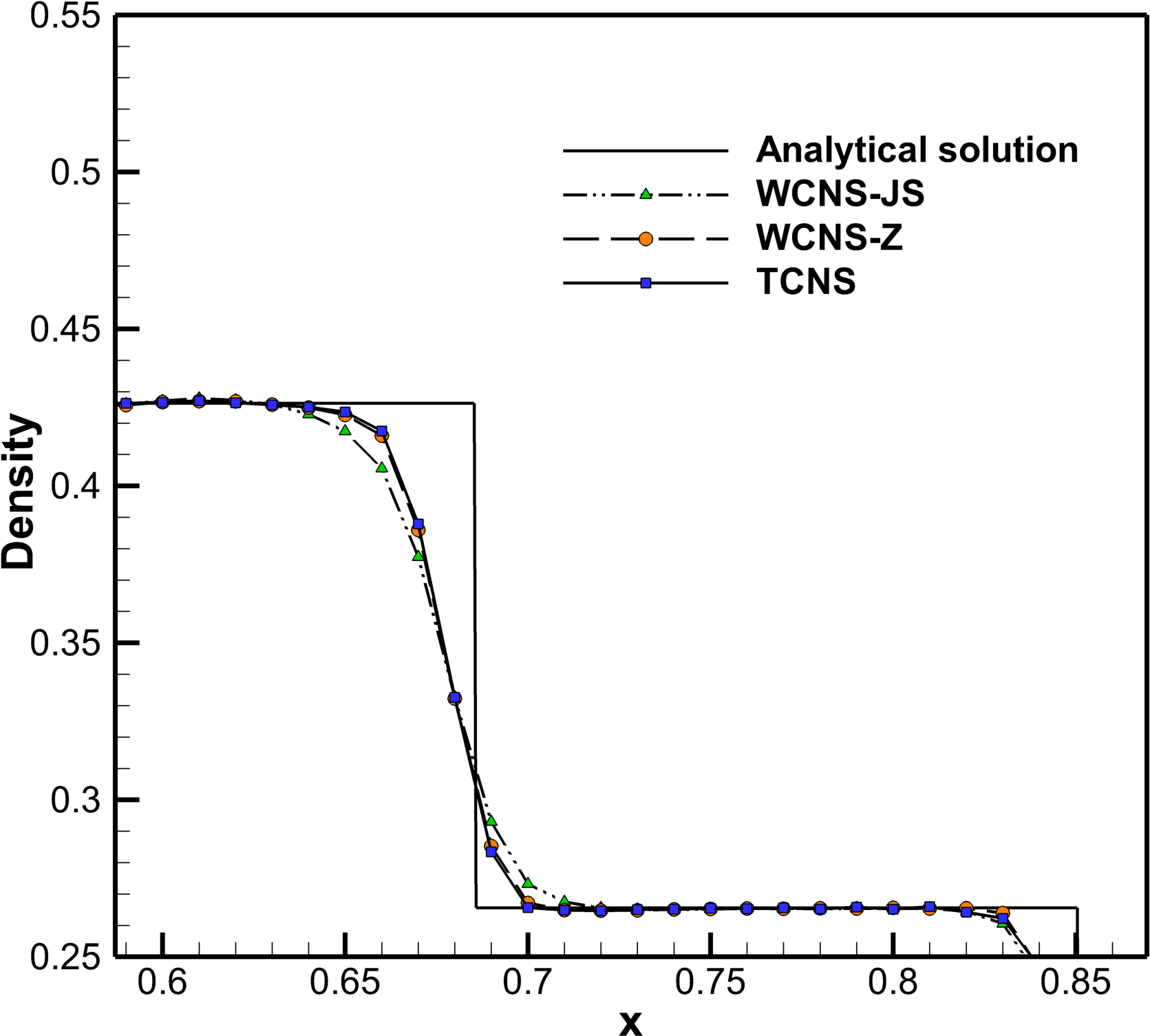}}}
\caption{\label{fig:sod-density} Sod problem: numerical solutions and the exact solution at $t=0.2$. } 
\end{center}
\end{figure}

Overall, good agreement of numerical solutions with the exact solution is achieved for all of the schemes shown in Fig.~\ref{fig:sod-density1}. Steep gradient solutions such as shock wave and contact discontinuity are captured without oscillations. In Fig.~\ref{fig:sod-density2}, WCNS-Z and TCNS schemes yield a relative sharp profile compared to WCNS-JS in the vicinity of contact discontinuity, and TCNS is found with a better performance of resolving the contact discontinuity when compared with WCNS-Z.

The initial conditions for the Lax shock-tube problem is
\begin{equation}
(\rho, u, p)=
\begin{cases}
\begin{matrix}
 (0.445, 0.698, 3.528)        &  \quad  x\in [0,0.5], \\
 (0.5,0,0.571)  & \quad  x\in  \left(0.5,1\right].
\end{matrix}
\end{cases}
\end{equation}
This case is run on a grid of $N=101$ points with uniform distribution, and the results at  $t=0.14$ are given.
\begin{figure}[h!t]
\begin{center}
\subfigure[\label{fig:lax-density1}{Density}]{
\resizebox*{5.5cm}{!}{\includegraphics{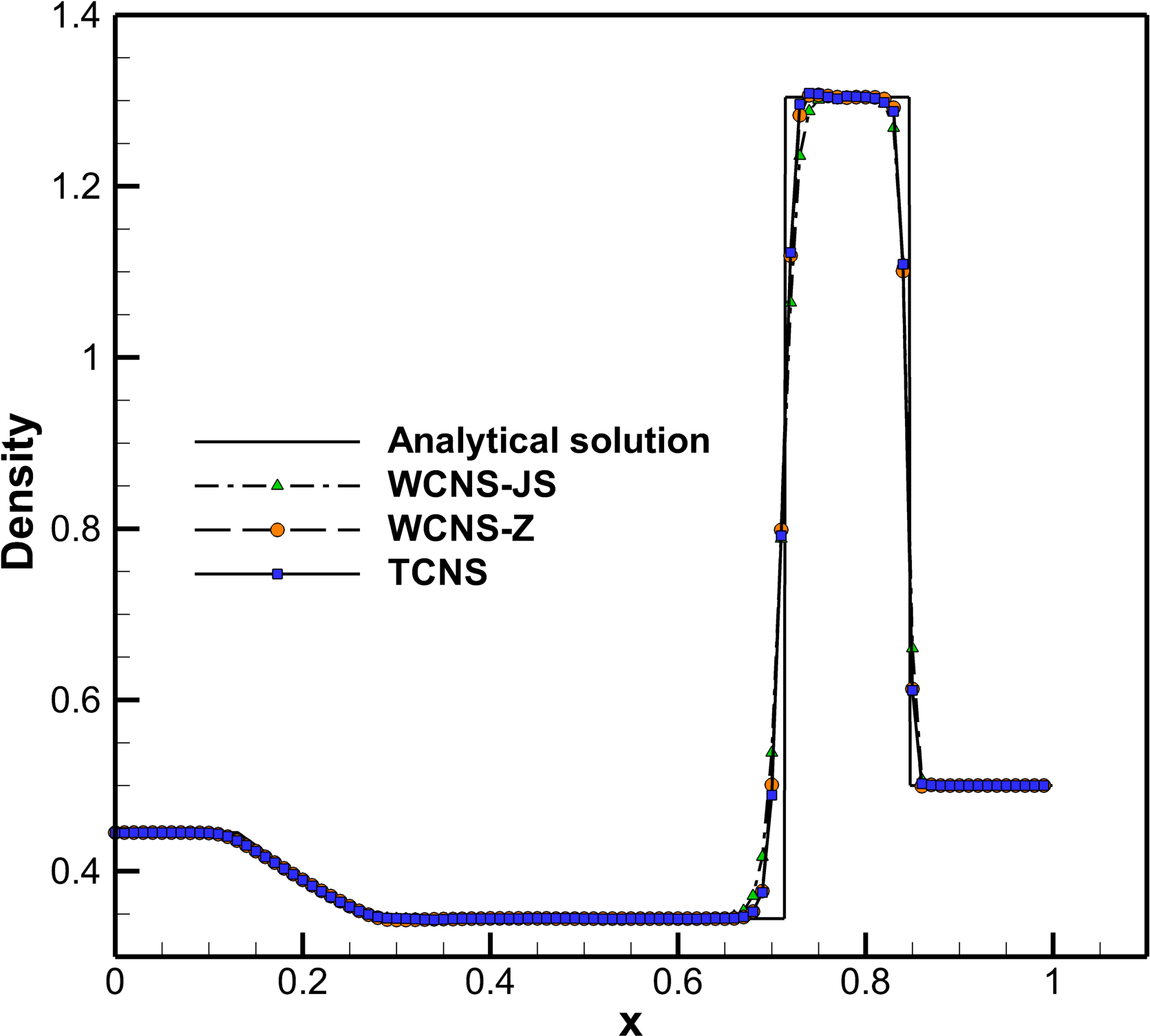}}}
\subfigure[\label{fig:lax-density2}{Velocity}]{
\resizebox*{5.5cm}{!}{\includegraphics{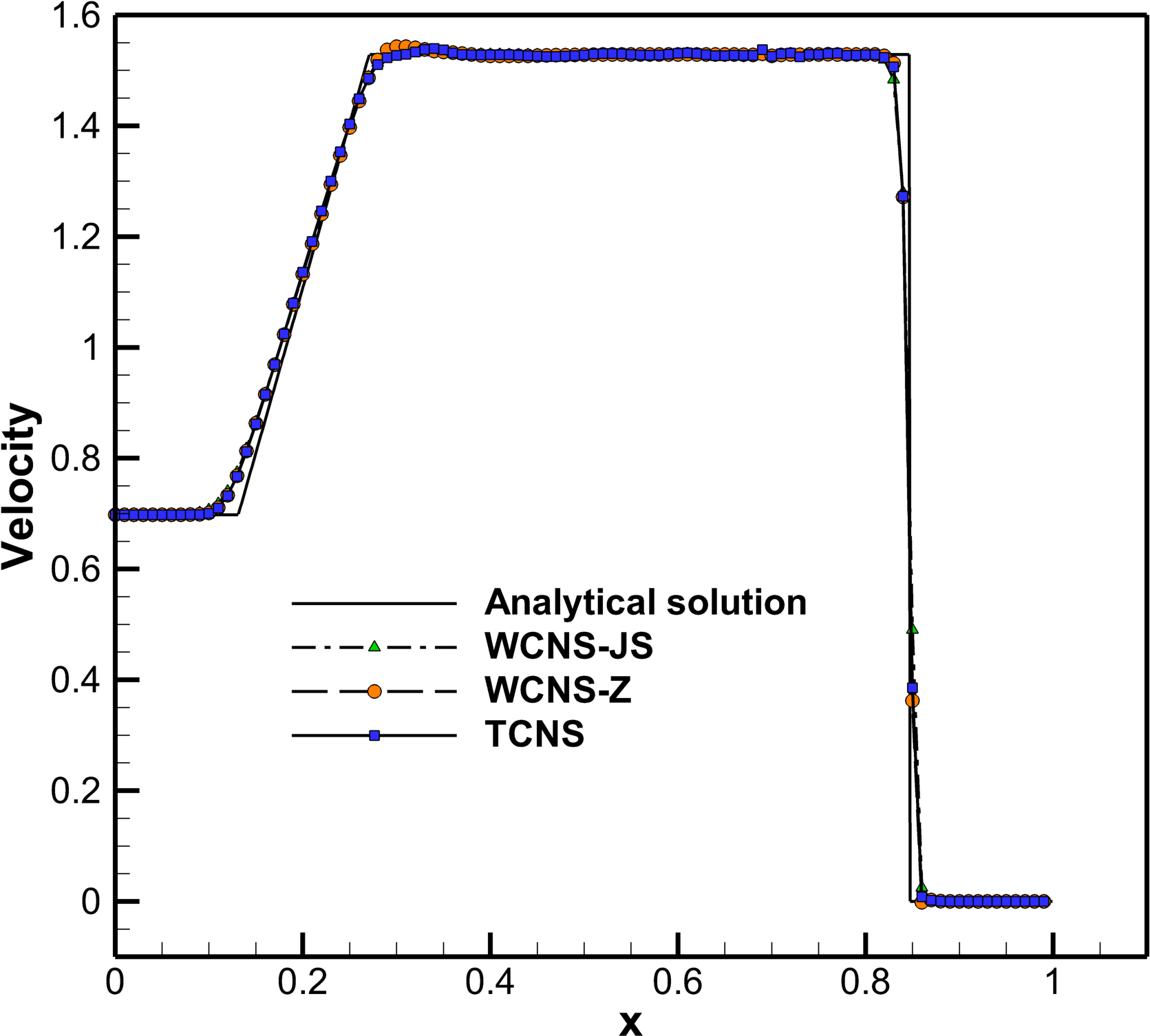}}}
\caption{\label{fig:lax-density} Lax problem: numerical solutions and the exact solution at $t=0.14$. }
\end{center}
\end{figure}

Density and velocity distributions are compared in Fig.\ref{fig:lax-density1}, and Fig.~\ref{fig:lax-density2} respectively.
WCNS-Z overshoots the velocity slightly at the tail of the expansion fan, but TCNS shows accurate and oscillation-free result. WCNS-JS smears the shock wave significantly, indicating extra numerical dissipation.


\subsection{Shock-density wave interaction}
The shock-density wave interaction problem \cite{SHU198932} is characterized by a right moving Mach 3 shock interacting with sine waves in  density field. The multi-scale wave structure is evolved after the shock wave interacts with the oscillating density wave, and both the shock-capturing and wave-resolution capabilities are evaluated thereafter.

The problem is initialized by
\begin{equation}
(\rho, u, p)=
\begin{cases}
\begin{matrix}
 (3.857, 2.629, 10.333),       &  \quad  x\in [0,1], \\
 \left(1 + 0.2\text{sin}(5x),0, 1\right),  & \quad  x\in  \left(1,10\right].
\end{matrix}
\end{cases}
\end{equation}
This case is run on a grid of $N=201$ points which are uniformly distributed and the final time is $t=1.8$. Numerical solution of WCNS-JS on a grid of $N=2001$ is used as the reference "exact" solution.

\begin{figure}[H]
\begin{center}
\subfigure[\label{fig:shu-density1}{}]{
\resizebox*{5.5cm}{!}{\includegraphics{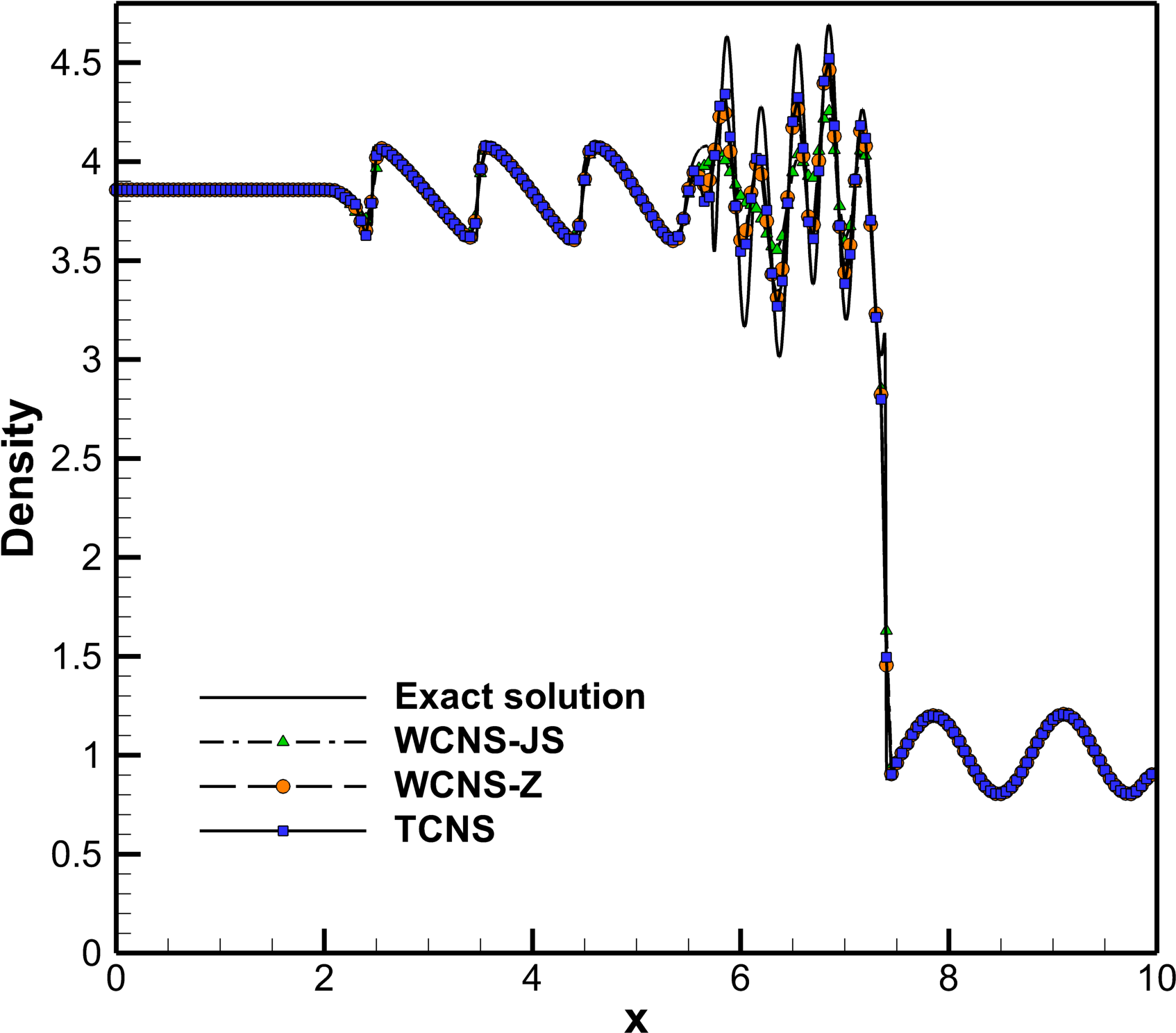}}}
\subfigure[\label{fig:shu-density2}{}]{
\resizebox*{5.5cm}{!}{\includegraphics{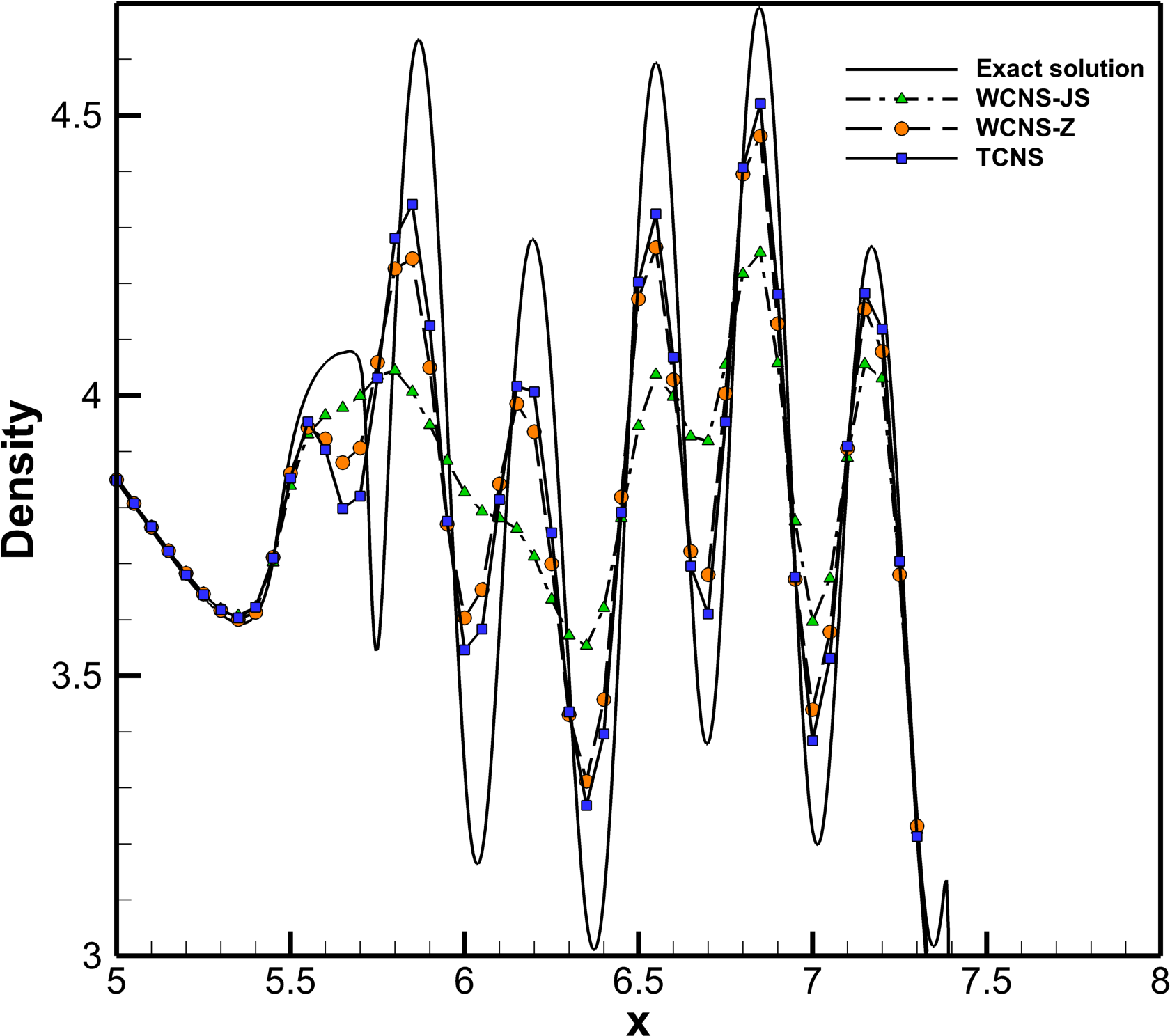}}}
\caption{\label{fig:shu-density} Shock-density wave interaction problem: numerical solutions and the exact solution at $t=1.8$.}
\end{center}
\end{figure}

As shown in Fig.~\ref{fig:shu-density}, TCNS produces considerably better resolved density waves behind the shock wave compared with WCNS-JS and WCNS-Z. Particularly, the result of WCNS-JS indicates strong numerical dissipation, since the small-scale wave structure is significantly smeared. Moreover, TCNS does not produce spurious oscillations in the flow field.

\subsection{Two-dimensional Riemann problem}

In this section, the configurations  3 and 6 out of 19 2-D Riemann problems used by Lax and Liu~\cite{Lax1998} are specifically taken to evaluate performances of the proposed numerical schemes.

\subsubsection{The configuration 3}
The initial conditions for the configuration 3 are given by
\begin{equation}
(\rho, u, v, p)=
\begin{cases}
\begin{matrix}
 (1.5, 0, 0, 1.5)         &  \quad  x\in \left[\frac{1}{2},1\right] \; \text{and}\; y\in \left[\frac{1}{2},1\right], \\
 (0.5323, 1.206, 0, 0.3)        & \quad   x\in \left[0 ,\frac{1}{2}\right) \; \text{and}\; y\in \left[\frac{1}{2}, 1\right], \\
 (0.138, 1.206, 1.206, 0.029)          &  \quad  x\in \left[0,\frac{1}{2}\right) \; \text{and}\; y\in \left[0,\frac{1}{2}\right), \\
 (0.5323, 0, 1.206, 0.3)         & \quad   x\in \left[\frac{1}{2},1\right] \; \text{and}\; y\in \left[0,\frac{1}{2}\right). \\
\end{matrix}
\end{cases}
\end{equation}
Boundary conditions are given by
\begin{equation}
\begin{aligned}
 \frac{\partial \mathbf{u}(t,x,y)}{\partial x} &= \mathbf{0}  , \quad   x= 0, 1, \quad \forall t, y, \\
 \frac{\partial \mathbf{u}(t,x,y)}{\partial y} &= \mathbf{0}  , \quad   y= 0, 1, \quad \forall t, x.
\end{aligned}
\end{equation}

\begin{figure}[H]
\begin{center}
\subfigure[\label{fig:2dR3-wcns}{WCNS-JS}]{
\resizebox*{5.5 cm}{!}{\includegraphics{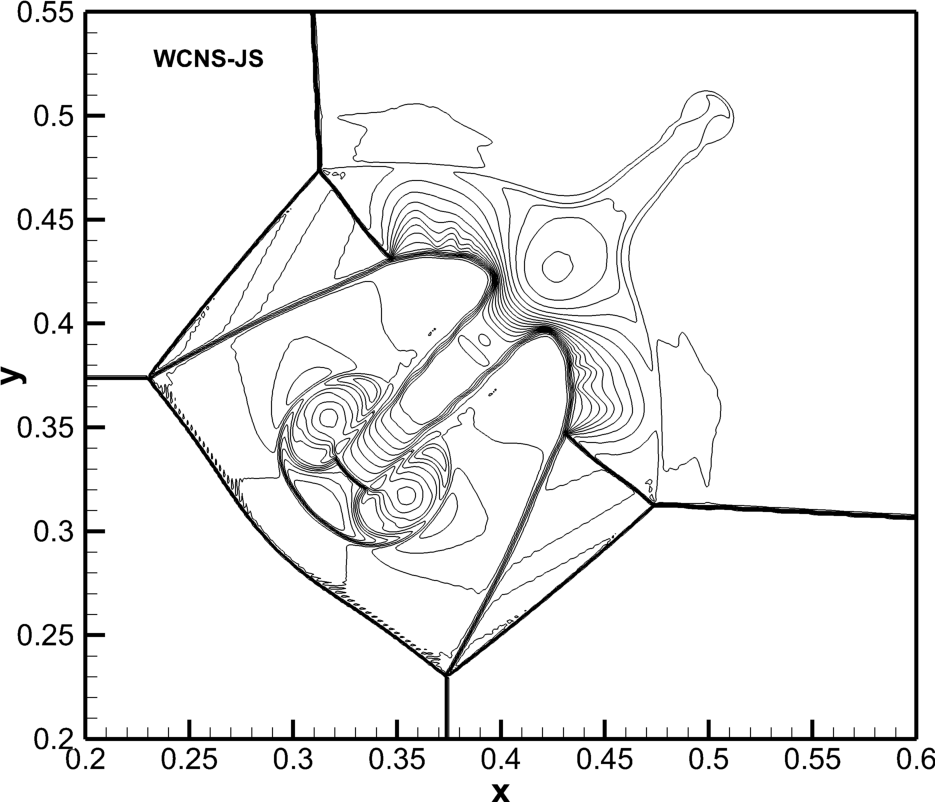}}}
\subfigure[\label{fig:2dR3-wcnsz}{WCNS-Z}]{
\resizebox*{5.5 cm}{!}{\includegraphics{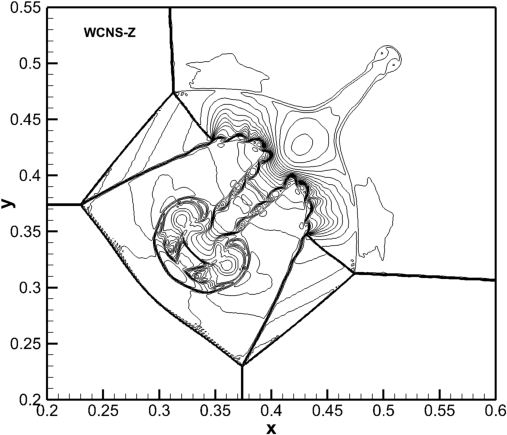}}}
\subfigure[\label{fig:2dR3-tcns}{TCNS}]{
\resizebox*{5.5 cm}{!}{\includegraphics{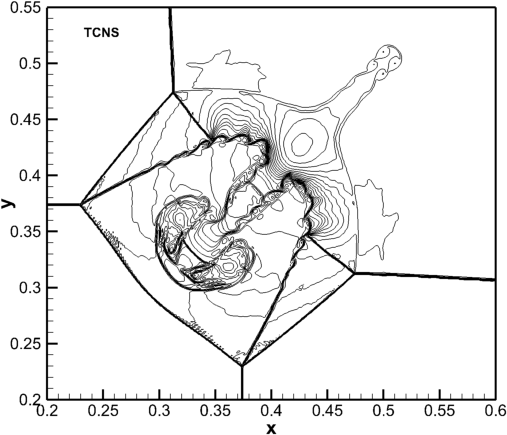}}}
\caption{\label{fig:2dR3} Configurations  3 of 2-D Riemann problems in \cite{Lax1998}: 30 density contours ranging from 0.1 to 1.8 at $t=0.23$ obtained  on a grid of $1024\times 2014$. }
\end{center}
\end{figure}

Figure~\ref{fig:2dR3} presents the results obtained by using WCNS-JS, WCNS-Z and TCNS schemes, at $t=0.3$. The density profiles are best predicted by TCNS. WCNS-Z yields a solution very close to that of TCNS, but TCNS captures more vortexes along the contact lines. WCNS-JS smears a few small scales that can be resolved by TCNS and WCNS-Z due to relative large dissipations.

\subsubsection{The configuration 6}

The initial conditions for the configuration 6 are given by
\begin{equation}
(\rho, u, v, p)=
\begin{cases}
\begin{matrix}
 (1, 0.75, -0.5, 1)         &  \quad  x\in \left[\frac{1}{2},1\right] \; \text{and}\; y\in \left[\frac{1}{2},1\right], \\
 (2, 0.75, 0.5, 1)        & \quad   x\in \left[0 ,\frac{1}{2}\right) \; \text{and}\; y\in \left[\frac{1}{2}, 1\right], \\
 (1, -0.75, 0.5, 1)          &  \quad  x\in \left[0,\frac{1}{2}\right) \; \text{and}\; y\in \left[0,\frac{1}{2}\right), \\
 (3, -0.75, -0.5, 1)         & \quad   x\in \left[\frac{1}{2},1\right] \; \text{and}\; y\in \left[0,\frac{1}{2}\right). \\
\end{matrix}
\end{cases}
\end{equation}
Boundary conditions are the same as in the last case.

\begin{figure}[H]
\begin{center}
\subfigure[\label{fig:2dR6-wcns}{WCNS-JS}]{
\resizebox*{5.5 cm}{!}{\includegraphics{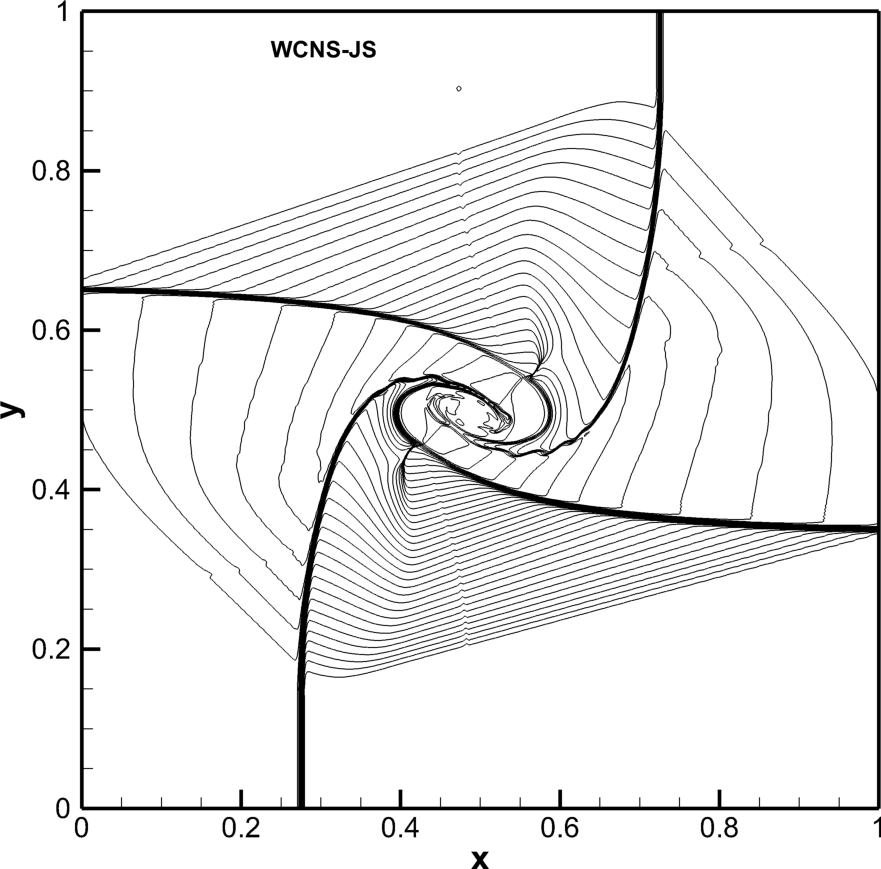}}}
\subfigure[\label{fig:2dR6-wcnsz}{WCNS-Z}]{
\resizebox*{5.5 cm}{!}{\includegraphics{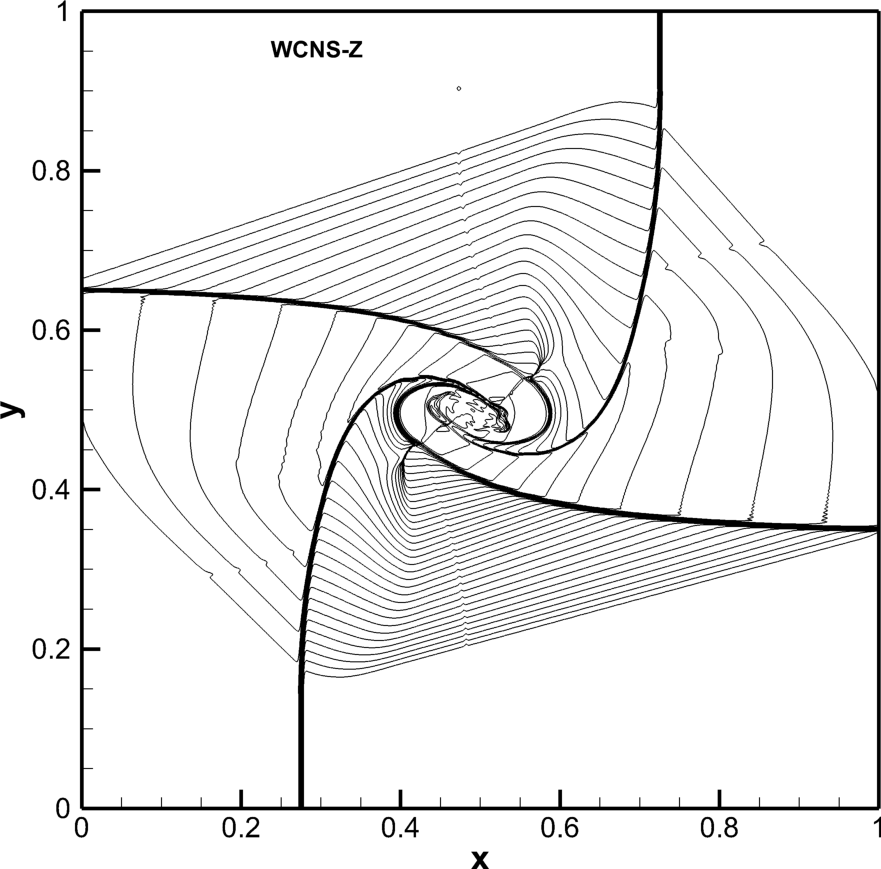}}}
\subfigure[\label{fig:2dR6-tcns}{TCNS}]{
\resizebox*{5.5 cm}{!}{\includegraphics{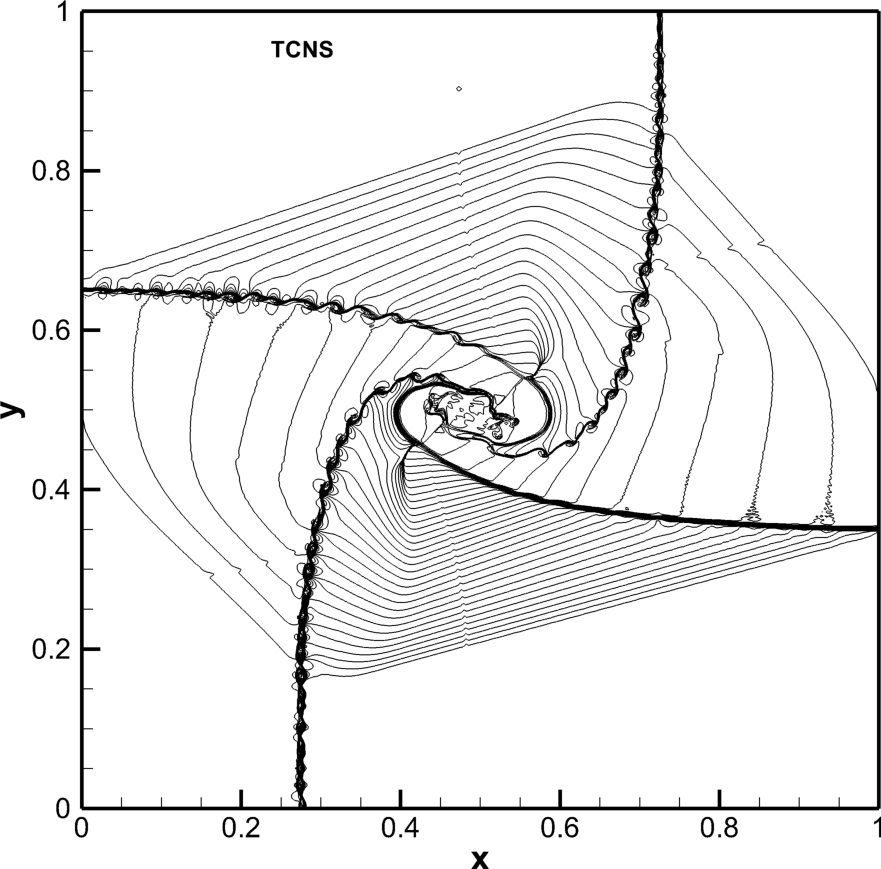}}}
\caption{\label{fig:2dR6} Configurations 6 of 2-D Riemann problems in \cite{Lax1998}: 40 density contours ranging from 0.1 to 2.9 at $t=0.3$ obtained on a grid of $1024\times 2014$. }
\end{center}
\end{figure}

Results obtained by using WCNS-JS, WCNS-Z and TCNS schemes at $t=0.3$ are shown in Fig.~\ref{fig:2dR6}.
The results of  WCNS-JS and WCNS-Z are similar. Whereas, using TCNS obtains abundant small scale flow structures along the contact lines, indicating lower numerical dissipation.

\subsection{Shock vortex interaction}
\begin{figure}[H]
\begin{center}
\subfigure[\label{fig:2ds-wcns}{WCNS-JS}]{
\resizebox*{5.5 cm}{!}{\includegraphics{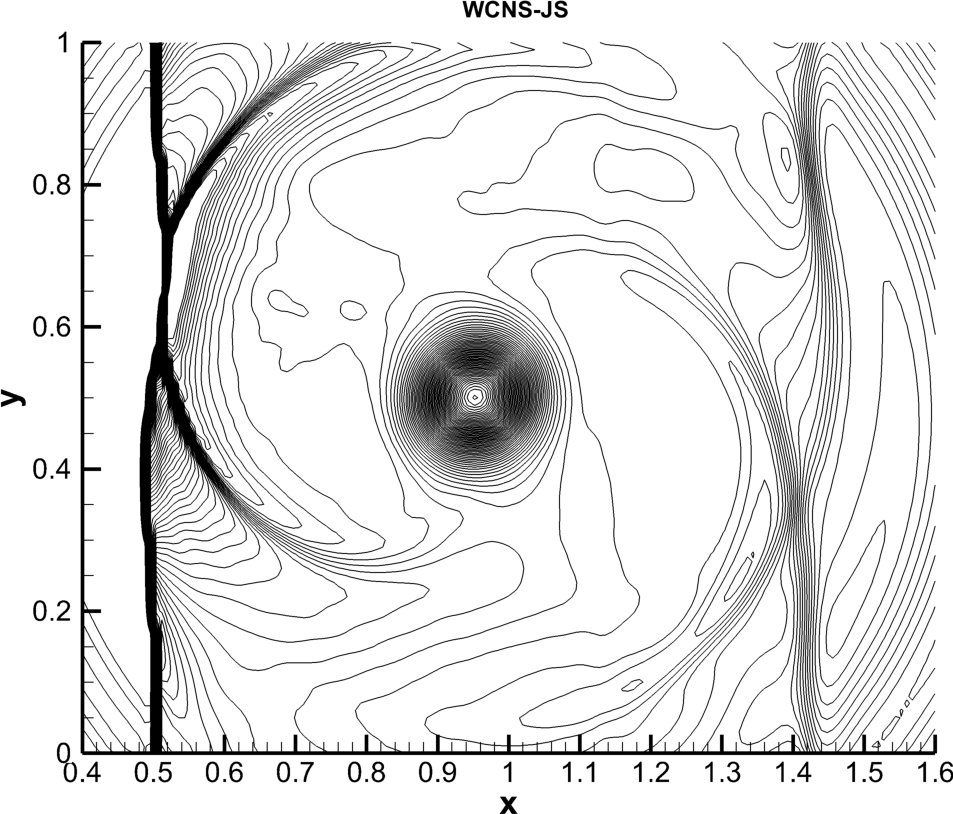}}}
\subfigure[\label{fig:2ds-wcnsz}{WCNS-Z}]{
\resizebox*{5.5 cm}{!}{\includegraphics{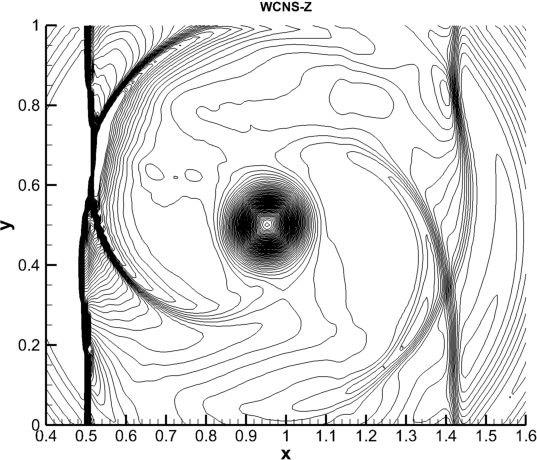}}}
\subfigure[\label{fig:2ds-tcns}{TCNS}]{
\resizebox*{5.5 cm}{!}{\includegraphics{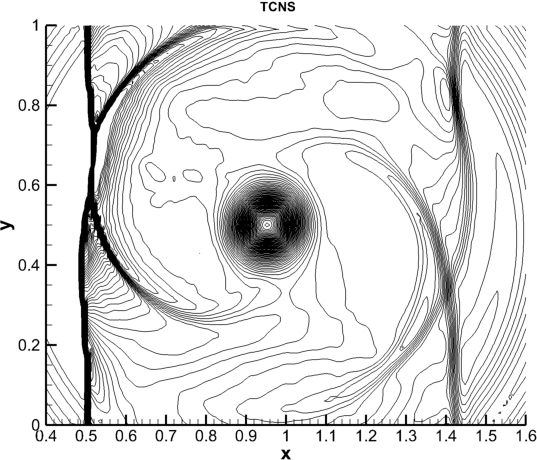}}}
\caption{\label{fig:2ds6} Shock vortex interaction problem: 90 density contours ranging from 1.19 to 1.37. }
\end{center}
\end{figure}

This problem was modeled in two dimensions by Jiang and Shu~\cite{Jiang1996}. It involves a vortex perturbing a stationary shock.
The computational domain is $[0,2] \times [0,1]$. This field is discretized by $251\times 101$ grid. Initially, a stationary shock is located at $x=0.5$ normal to the $x$ direction. The left state of this shock is specified as $(\rho, u, v, p) = (1, 1.1\sqrt{\gamma}, 0, 1)$. A small vortex centered at (0.25, 0.5) is superposed to the flowfiled on the left hand side of the normal shock. The superposition is performed through perturbations on velocity $(u,v)$, temperature $T$, given by $T=p/\rho$, and entropy $S$, defined as $S=\text{ln} (p/\rho^{\gamma})$ of the mean flow. Specifically, the perturbation variables are

\begin{equation}
\begin{cases}
 &\tilde{u} = \epsilon \tau e^{a (1-\tau^2)} \text{sin}\theta \\
 &\tilde{u} =- \epsilon \tau e^{a (1-\tau^2)} \text{cos}\theta  \\
 &\tilde{T} = - \frac{(\gamma -1 ) \epsilon^2 e^{2a (1-\gamma^2)} }{4 a \gamma} \\
 &\tilde{S} = 0                              \\
\end{cases}
\end{equation}
where $\tau = r/r_c$, and $r=\sqrt{{(x-x_c)^2} + {(y-y_c)^2} }$, $r_c=0.05$, $\epsilon = 0.3$, $a =0.204$ are taken from the reference~\cite{Jiang1996}. The results at $t=0.6$ are presented in Fig.\ref{fig:2ds6}.

\begin{figure}[h!t]
 \centering
 \includegraphics[width=8cm]{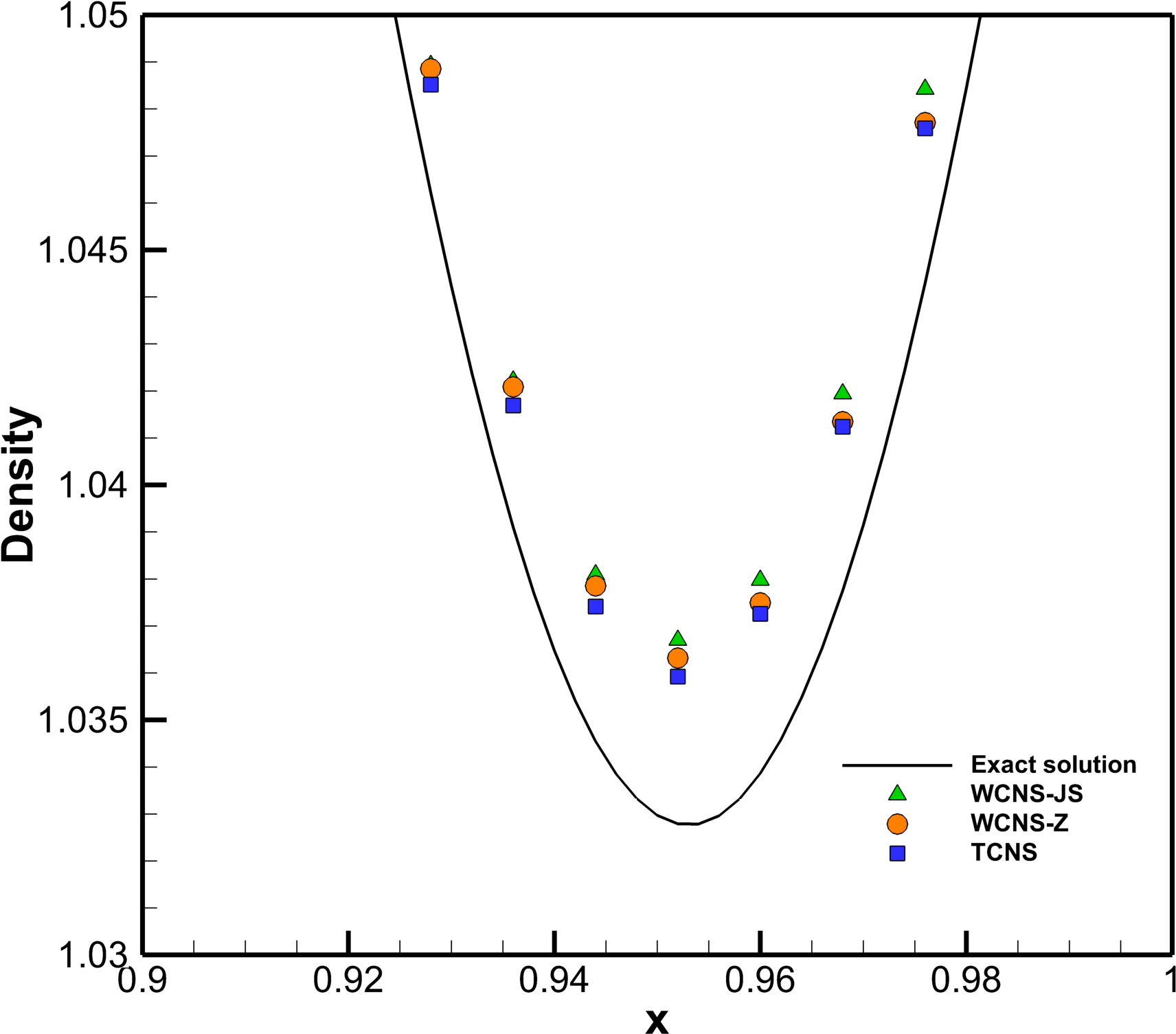}
 \caption{ \label{fig:shock-vortex-density}
 Shock vortex interaction problem: density distribution along $y = 0.5$.}
\end{figure}

The flow contours of different schemes are similar.
The density distribution along the central line of the flow field are further shown in Fig.\ref{fig:shock-vortex-density}, where the result obtained by using WCNS-Z on a refined mesh of $1001\times 401$ is used as the ``exact" solution. It can be found that the captured vortex of TCNS is more accurate, comparing with the other two results of WCNS-JS and WCNS-Z.

\subsection{Rayleigh-Taylor instability}
\begin{figure}[H]
\begin{center}
\subfigure[\label{fig:rt-1}{Grid resolution $128\times512$ }]{
\resizebox*{12 cm}{!}{\includegraphics{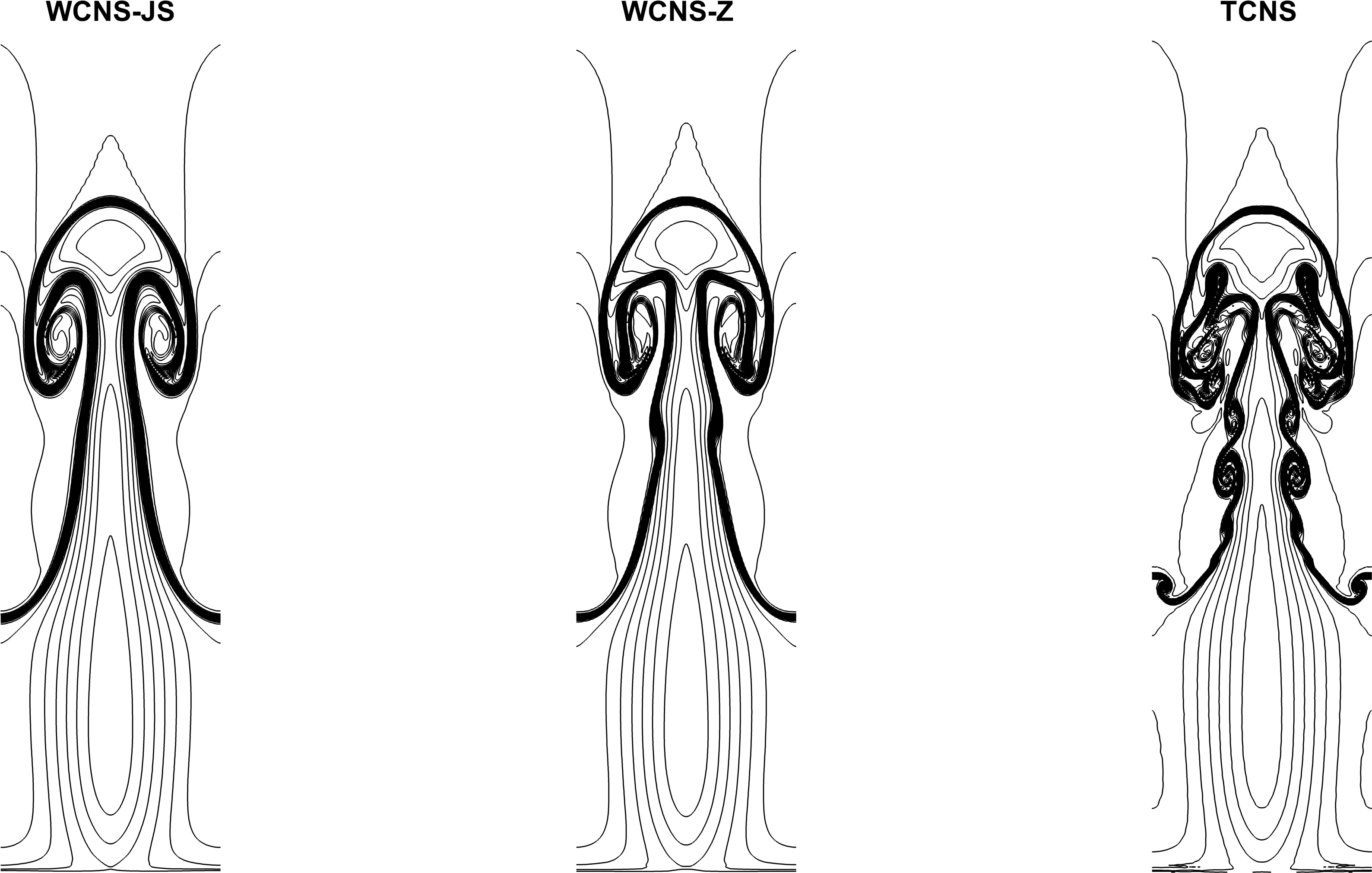}}}
\subfigure[\label{fig:rt-2}{Grid resolution $256\times 1024$}]{
\resizebox*{12cm}{!}{\includegraphics{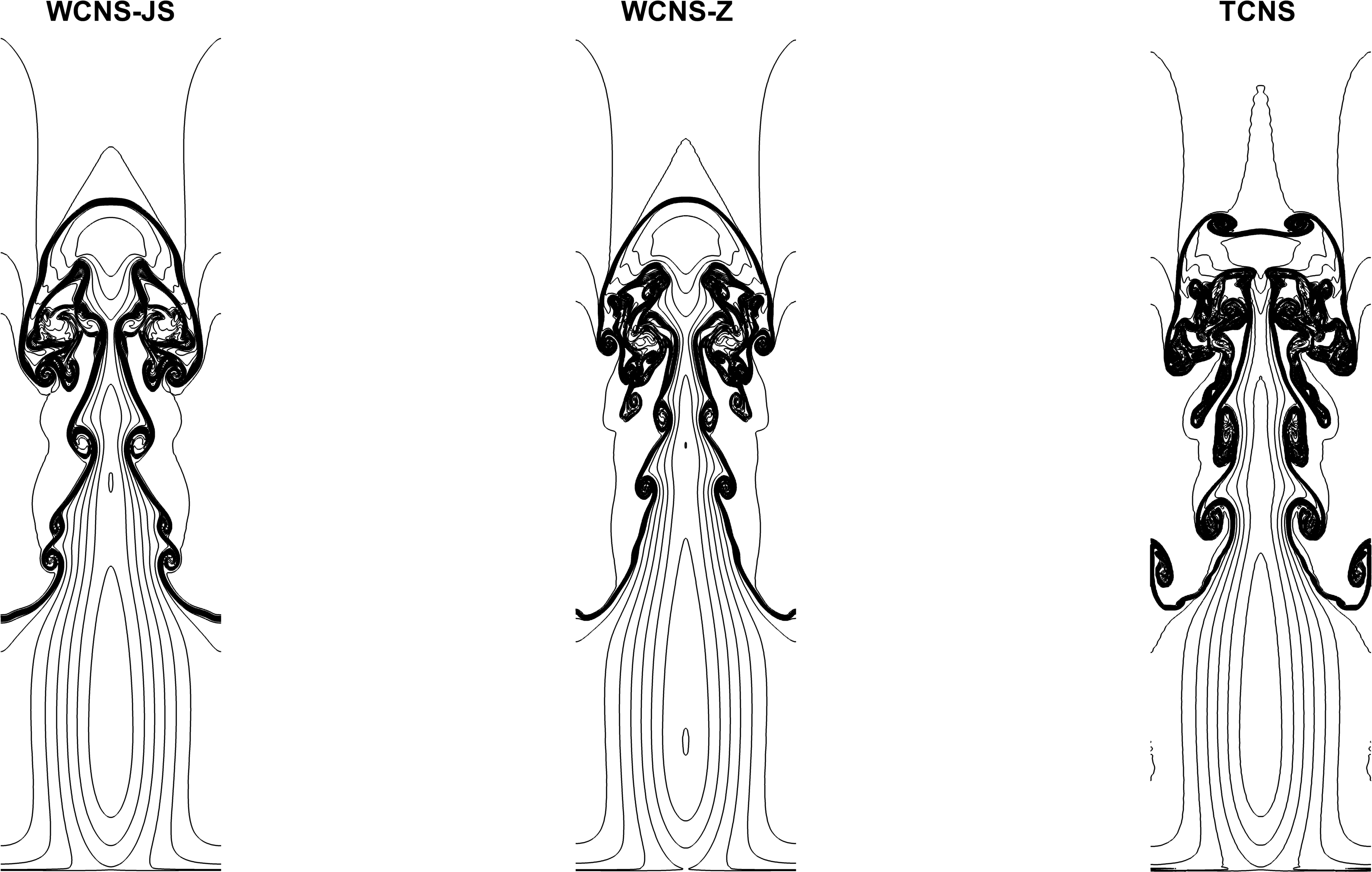}}}
\caption{\label{fig:rt-density}Rayleigh-Taylor instability problem: 30 density contours ranging from 0.9 to 2.2. }
\end{center}
\end{figure}

Rayleigh-Taylor instability problem  which  contains   both  discontinuities   and   complex   flow  structures \cite{XU2005458}, is also used to examine the numerical dissipation of the presented method. The initial conditions are given by
\begin{equation}
(\rho, u, v, p)=
\begin{cases}
\begin{matrix}
 (2, 0, -0.025 a\; \text{cos}(8\pi x), 1+2y)          &  \quad   x \in \left[0 , 0.25 \right]\;  \text{and}\; y\in \left[0 ,0.5\right), \\
 (1, 0, -0.025 a\; \text{cos}(8\pi x), 1+3/2)         &  \quad   x \in \left[0 , 0.25 \right] \;  \text{and}\; y\in \left[0.5 ,1 \right],
\end{matrix}
\end{cases}
\end{equation}
where $a$ is the speed of sound, given by $a=\sqrt{\gamma \frac{p}{\rho} }$ and a different $\gamma = \frac{5}{3}$ is used for this specific case. Reflecting boundary conditions are imposed at the left and right side of the domain, and constant boundary conditions are given for the top and the bottom sides, in details
\begin{equation}
(\rho, u, v, p)=
\begin{cases}
\begin{matrix}
 (1, 0, 0, 2.5)         &  \quad   y= 1, \quad \forall t, x, \\
(2, 0, 0, 1)            &  \quad   y= 0, \quad \forall t, x.
\end{matrix}
\end{cases}
\end{equation}
Two source terms $\rho$, and $\rho v$ are added to the right hand side of the third and the fourth equation, respectively. Two sets of grids are used, i.e. $128\times 512$ and $256\times 1024$. Density profiles at final time $t=1.95$ calculated by WCNS-JS, WCNS-Z and TCNS   are shown in Fig.~\ref{fig:rt-density}.

TCNS has resolved much more abundant vortical  structures  than  WCNS-JS  and   WCNS-Z, suggesting that TCNS is significantly less dissipative.  Furthermore,  the   low-dissipation
property of TCNS induces symmetry breaking  of the  flow  field.
In particular, the small structures resolved by using TCNS  with grid resolution   of $128 \times 512$ are  comparable to that of using  WCNS-JS with grid resolution   of   $256 \times 1024$. Even the result of WCNS-Z only shows little advantage (if any) while doubling the grid resolution.
\section{Conclusions} \label{sec:Conclusions}

As a subsequent work of compact nonlinear scheme and weighed compact nonlinear scheme,   we  have  introduced  a  novel compact nonlinear scheme in this article, by applying the ENO-like stencil-selection procedure. The method is targeted optimal linear interpolation in the node-to-midpoint interpolation step of the compact nonlinear scheme, and thus the method is named as targeted CNS (TCNS).
Numerical results of one-dimensional scalar equations and one- or two-dimensional Euler equations are given to examine the performance of the method, involving  strong   discontinuities and broadband fluctuations.

The presented method uses a novel smoothness measurement which possesses strong scale  separation mechanism, and the ENO-like stencil-selection procedure is capable to target optimal linear weights while maintaining excellent shock-capturing capability. ADR analysis shows that TCNS recovers the background linear scheme up to higher wave number, even using a relative larger $C_T$.
Examining the numerical results, it can be found that TCNS has recovered optimal linear scheme in smooth field, where the optimal linear weights $d_k$ is directly applied in the node-to-midpoint interpolation. Whereas, WCNSs, including WCNS-JS and WCNS-Z, approximate the optimal linear weights but can not perfectly recover them. Therefore, the presented TCNS is capable to capture much more abundant wave structures in the simulations. Especially in the Rayleigh-Taylor instability problem, TCNS achieves similar or even better result comparing with WCNS-JS on a coarser grid.

\section*{Acknowledgments}
 Financial support for this work was provided through grant number 20187413071020008.

\bibliography{ref}

\end{document}